\documentclass{mn2e}
\usepackage{times,psfig}

\begin{document}
\hsize=6truein

\renewcommand{\thefootnote}{\fnsymbol{footnote}}
\newcommand{\chandra}{{\it Chandra} }
\newcommand{\asca}{{\it ASCA} }
\newcommand{\pspc}{{\it Rosat} PSPC }
\newcommand{\xmm}{{\it XMM-Newton} }

\title[\chandra observations of A1795]
{Deep inside the core of Abell~1795: the \chandra view}

\author[S. Ettori et al.]
{\parbox[]{6.in} {S. Ettori$^{1,2}$,
A.C. Fabian$^1$, S.W. Allen$^1$ and R.M. Johnstone$^1$ \\
\footnotesize
$^1$ Institute of Astronomy, Madingley Road, CB3 0HA Cambridge \\ 
$^2$ ESO, Karl-Schwarzschild-Str. 2, D-85748 Garching, Germany
}}                                            
\date{Resubmitted 2001 November 23. Submitted 2001 May 9.}
\maketitle

\begin{abstract}
We present X-ray spatial and spectral analysis of the \chandra data
from the central 400 $h_{50}^{-1}$ kpc of the cluster of
galaxies Abell~1795.  
The plasma temperature rises outwards by a factor of 3, whereas the
iron abundance decreases by a factor of 4. The spatial distribution of
Oxygen, Neon, Sulphur, Silicon and Iron shows that supernovae Type Ia
dominate the metal enrichment process of the cluster plasma within the
inner 150 kpc. Resolving both the gas density and temperature in 9
radial bins, we recover the gravitational mass density profile and
show that it flattens within 100 kpc as $\rho_{\rm DM} \propto
r^{-0.6}$ with a power law index flatter than --1 at $>3
\sigma$ level. The observed motion of the central galaxy and the
presence of excesses and deficits along the North-South direction in
the brightness distribution indicate that the central cluster 
region is not relaxed. 
In the absence of any non-gravitational heating source, 
the data from the inner $\sim$ 200 kpc indicate the presence 
of a cooling flow with an integrated 
mass deposition rate of about 100 $M_{\odot}$ yr$^{-1}$.
Over the same cluster region, the observed rate of 74 $M_{\odot}$ yr$^{-1}$
is consistent with the recent \xmm
Reflection Grating Spectrometer limit of 150 $M_{\odot}$ yr$^{-1}$.

\end{abstract}


\section{INTRODUCTION} 
Abell 1795 (Abell, Corwin \& Olowin 1989)
is a nearby rich cD galaxy cluster well studied at optical, 
radio and X-ray wavelengths. In X-rays, it shows a relaxed structure
(Buote \& Tsai 1996) and a surface brightness strongly peaked in the
centre, where cooler gas is present (Fabian 1994, Briel \& Henry 1996, 
Allen \& Fabian 1997, Ikebe et al. 1999, Allen et al. 2001, 
Tamura et al. 2001).  
Strong optical emission lines around the cD galaxy (Cowie et al. 1983), 
an excess in the blue optical light (Johnstone, Fabian \& Nulsen 1987; 
Cardiel, Gorgas \& Aragon-Salamanca 1997) and emission in ultra-violet
(Mittaz et al. 2001) due to the formation of massive
stars, perhaps triggered by the central radio source 4C~26.42
(McNamara et al. 1996), support a scenario in which
the central peak of cooler gas is due to a massive cooling flow 
(Edge et al. 1992, Fabian 1994).

In a previous paper (Fabian et al. 2001b), we have presented a
spatial analysis in three X-ray colors of the \chandra observation
of the core of A1795.
In the present work, we analyze the spectral characteristics 
of the intracluster medium, studying the metal enrichment, 
deprojecting the physical properties and determining the form 
of the gravitational potential within the cluster core.

Hereafter, we assume the values of the cosmological parameters to be
$H_0 = 50$ km s$^{-1}$ Mpc$^{-1}$, $\Omega_{\rm m} = 1$, 
$\Omega_{\Lambda} = 0$.
At the nominal redshift of A1795 ($z=0.0632$, Girardi et al. 1998),
an angular scale of 1 arcsec ($\sim 2$ ACIS pixels) corresponds
to a physical length of 1.65 kpc.
For the currently popular cosmology of $(H_0,\Omega_{\rm m},
\Omega_{\Lambda}) =(70, 0.3, 0.7)$, distances change by a factor of 0.74, 
luminosity by 0.74$^2$ = 0.54 and the age of the Universe at the cluster
redshift by 1.06.

All the errors quoted are $1 \sigma$ (68.3 per cent level of confidence)
unless stated otherwise.

\section{\chandra datasets}
 
In Table~1, we present a summary of the \chandra (Weisskopf et al. 2000)
observations which were both made with the back-illuminated CCD S3 of the 
Advanced Camera for Imaging and Spectroscopy (ACIS) instrument.
We have used the CIAO software (version~2.1) 
to clean further the revision-1 
level-2 events files provided by the standard pipeline processing that
filters for \asca grade classifications of 0,2,3,4,6 and defines 
Good Time Intervals for a nominal total exposure given in parenthesis 
in Table~1.
The FP--120\degr dataset has been analyzed with the August 2001 release
of of ACIS response products (Fits Encoded Function and gain files;
see at {\tt http://asc.harvard.edu/cal/Links/Acis/acis/}
{\tt Cal\_prods/matrix/matrix.html}), whereas the FP--110\degr events
file is calibrated according to the previous release of December 1999.
Therefore the original FP--120\degr dataset has been reprocessed 
with {\it acis\_process\_events} routine in CIAO 2.1.3
with the gain files available in CALDB 2.7.
We have also corrected for time intervals with (i) bad attitude 
solutions, and (ii) high background with flares in the lightcurve 
with relative value larger than 20 per cent the mean count rate.
This process reduces the effective exposure by about 
5 per cent in the two datasets.

\begin{table}
\caption[]{\chandra observations summary of A1795. ``BI" stands for 
``Back Illuminated CCD". The temperature to which the Focal Plane 
(FP) is cooled is given for each exposure. 
As discussed in the text, this temperature
is relevant for the definition of the calibration files.
}
\begin{tabular}{cccc}
Chip  &  Date & Exposure (sec) \\
BI--S3, FP--110\degr & 20 Dec 1999 & 18370 (19522) \\
BI--S3, FP--120\degr & 21 Mar 2000 & 18927 (19627) 
\end{tabular}
\end{table}

\section{Imaging analysis}

\begin{figure}
\psfig{figure=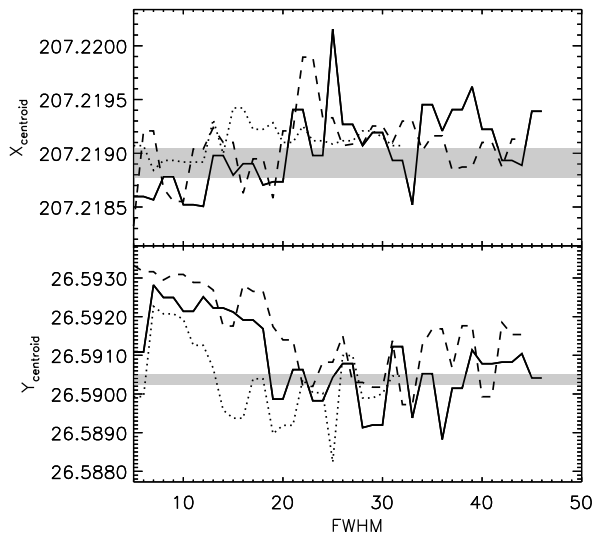,width=.5\textwidth}
\caption[]{X-ray centre (shaded region with a width of 1 arcsec) and
centroids estimated within a box with half-width equal to 
0.637 $\times$ FWHM in units of arcsec
(routine {\tt CNTRD} in Interactive Data Language)
and in three different X-ray colour images (0.5 -- 0.8 keV band: {\it 
solid line}; 0.8 -- 1.5 keV band: {\it dotted line}; 
1.5 -- 7 keV band: {\it dashed line}). 
} \label{fig:center}
\end{figure}

\begin{figure}
\psfig{figure=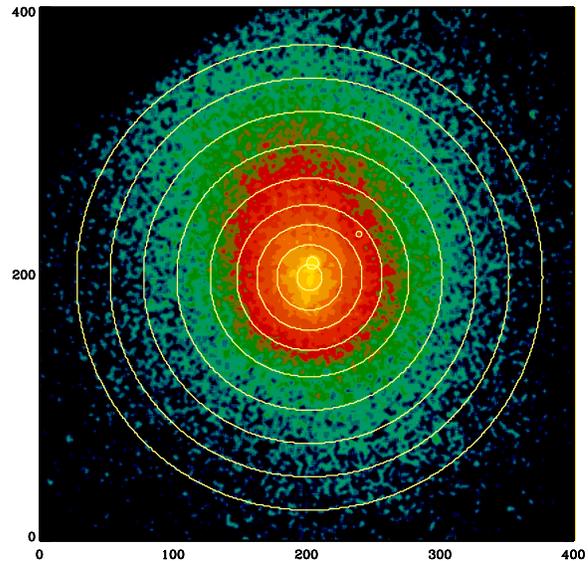,width=.5\textwidth}
\caption[]{Annuli adopted in the spectral analysis overlaid
to a smoothed image of A1795 (units in pixel = 0.984 arcsec).
} \label{fig:ann}
\end{figure}

The optical position of the cD (which hosts the radio source 4C+26.42)
is (RA, Dec) 13$^h$ 48$^m$ 52\fs43, 26\degr 35\arcmin 34\farcs0 (J2000).
This is within the $5 \times 5$ arcsec$^2$ bright envelope to the north
of the X-ray emitting wake, also known as an H$\alpha$ emitter,
discussed in our previous work (Fabian et al. 2001b).
The X-ray centre for our analysis is chosen at coordinates
(RA, Dec) 13$^h$ 48$^m$ 52\fs54, 26\degr 35\arcmin 25\farcs3 (J2000).
This is consistent with the centroid of the X-ray emission
determined in three different X-ray colour images.
The behaviour of the centroid and the chosen centre are
plotted in Fig.~\ref{fig:center}.

The choice of the centre is also determined to make circular
annuli a good representation of the cluster surface brightness
as shown in Fig.~\ref{fig:ann}.
This is relevant in the deprojection analysis applied in the
present work that assumes a spherical symmetric emission projected
on the sky in circular regions.

\begin{figure*}
\hbox{
\psfig{figure=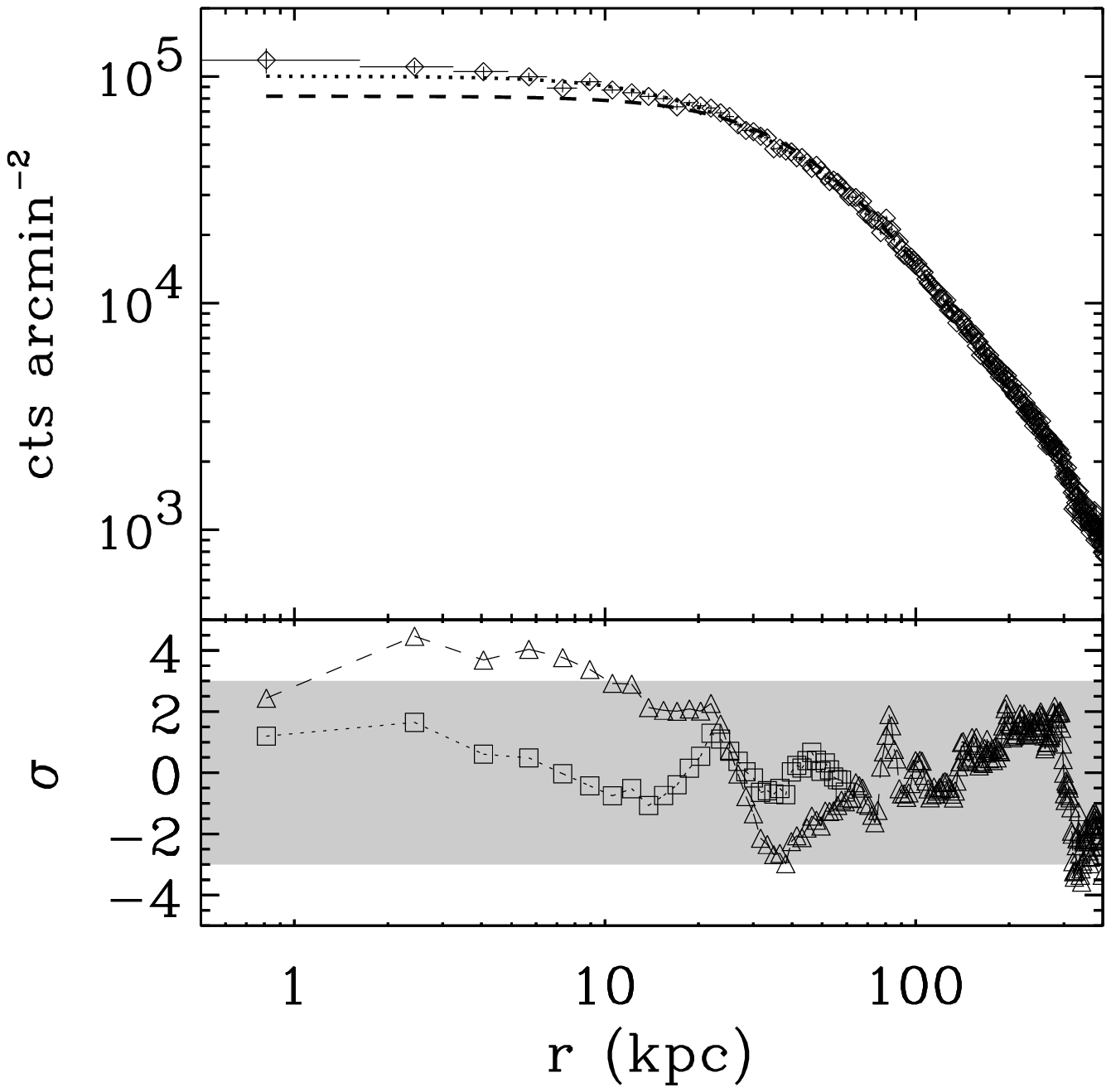,width=.5\textwidth}
\psfig{figure=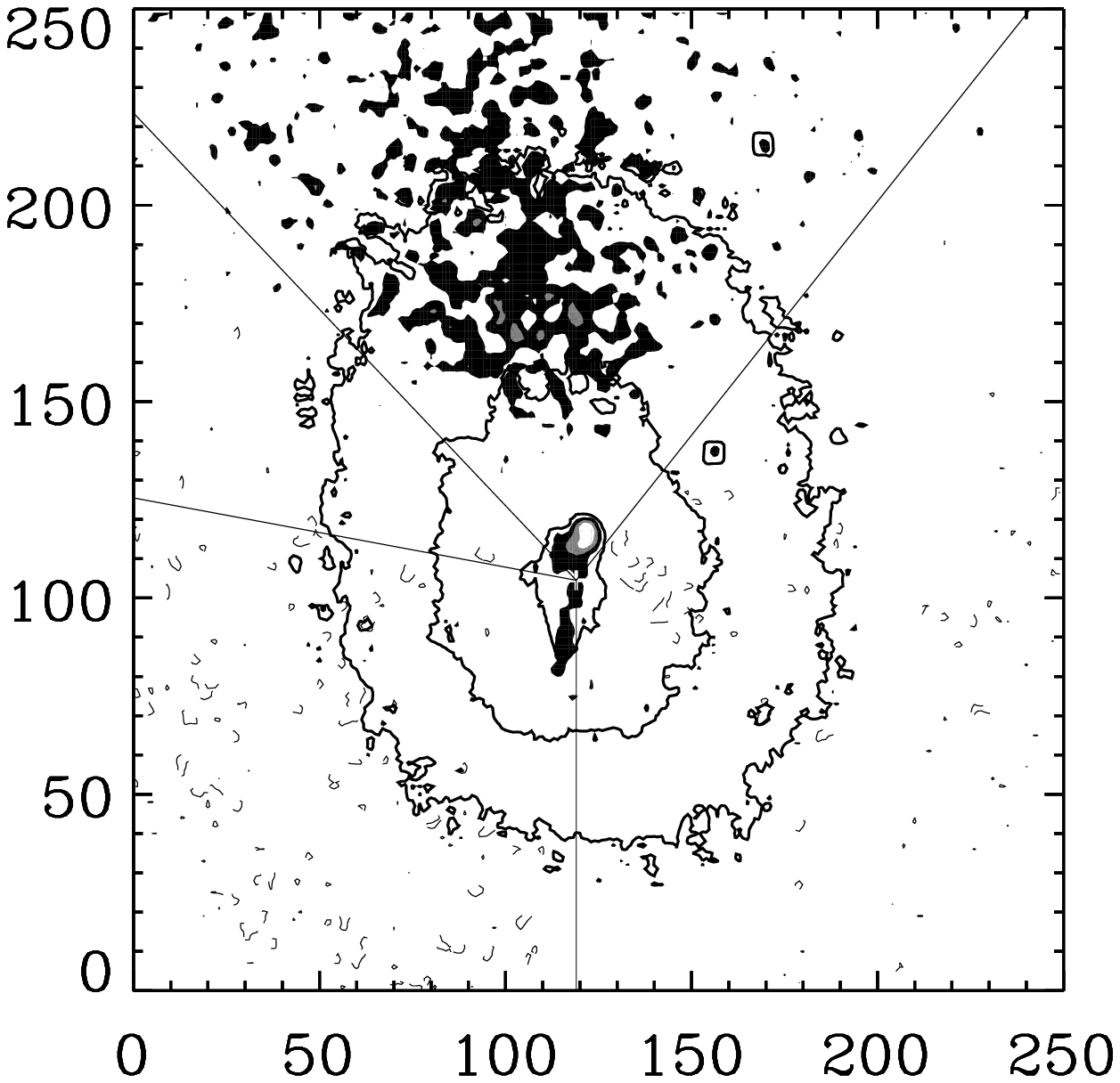,width=.5\textwidth}
} \caption[]{
(Left) The azimuthally averaged brightness profile 
after background subtraction and exposure correction is shown. 
Overplotted are two $\beta-$model, one fitted on 0--60 kpc (dotted line and 
squares), the other over the 0--400 kpc radial range (dashed line and 
triangles). The differences in $\sigma$ with respect to the model are 
smoothed over 5 bins for sake of clarity. Two main breaks are evident,
at about 25 and 300 kpc, respectively.
The shaded region indicates deviations within $\pm 3\sigma$.
(Right) $250 \times 250$ (1\arcsec-bin)$^2$ ($\sim 410 \times 410$ kpc$^2$)
map, smoothed on a 3 arcsec scale, of the differences between the original
exposure corrected image and the two-dimensional distribution of a 
$\beta-$model fitted over the 0--400 kpc radial range (dashed line 
in the panel on the left). The shaded regions indicate positive values 
of $(I-M)/\sqrt{M}$, whereas dotted contours shows negative values.
These deviations have an absolute value that ranges between 1 and 3. 
The lines that depart from the cross delimit the sectors in exam 
in Fig.~\ref{fig:sb_sect}.
} \label{fig:sbd}
\end{figure*}

The azimuthally averaged brightness profile is shown in Fig.~\ref{fig:sbd}.
Two main breaks are evident at about 25 and 300 kpc, respectively,
when the profile is compared with a smoothed distribution
provided, e.g., from a $\beta-$model (e.g. Cavaliere \&
Fusco-Femiano 1976) fitted over the radial range 0--400 kpc
(with best-fit parameters $r_{\rm c} = 49 \pm 1$ kpc, 
$\beta = 0.510 \pm 0.002$, $\chi^2$ = 971, 243 degrees of freedom). 
The presence of breaks in the brightness distribution is now
routinely observed in \chandra data of galaxy clusters
(e.g. Allen, Ettori, Fabian 2001, Markevitch et al. 2001, Vikhlinin
et al. 2001, Mazzotta et al. 2001) and is mainly associated with 
discontinuities in the gas density profile probably due either
to the aggregation of subclumps or to the ``sloshing" of the gas decoupled 
from dark matter into the changing gravitational potential 
(as suggested for the break in the surface brightness about 70\arcsec 
southward the central galaxy in A1795 from Markevitch, Vikhlinin \&
Mazzotta 2001).

It is worth noting that a $\beta-$model is generally 
not a good representation of the brightness profile of 
cooling flow clusters, like A1795, over large radial ranges
due to the peaked central emission (but see improvements
in modelling this kind of profiles obtained from 
\pspc observations in Mohr et al. 1999 and Ettori 2000).
When the central 200 kpc radius region is not included in the fit,
Ettori \& Fabian (1999) measure from a \pspc observation that extends
up to 1.5 Mpc a core radius of 250 kpc and a $\beta$ of 0.75, 
consistent with results from fits that include a 
model for the central excess (Mohr et al. 1999, Ettori 2000).
Due to the strong positive correlation between 
core radius and $\beta$ parameter, the evidence of a smaller
core in the brightness profile as observed from \chandra
implies a lower value of $\beta$ as estimated. 
Moreover, it is interesting to note the dependence of these 
parameters from the radial range considered. 
For example, we find that, from a statistical point of view, 
the best modelling with a $\beta-$model in the central 400 kpc of A1795
is obtained over the range 0--60 kpc 
($\chi^2$/dof = 45/34, probability P=0.09), with best-fit parameters
$r_{\rm c} = 23 \pm 2$ kpc and $\beta = 0.353 \pm 0.011$ that show 
a significant lower value of $r_{\rm c}$ and a corresponding 
lower value of $\beta$. We remind, however, that the $\beta-$model
is a convenient fitting function that 
assumes both an isothermal gas and the hydrostatic equilibrium
between the galaxy population, the intracluster plasma and the 
underlying gravitational potential. In the cooling flow
regions, the assumption of isothermality is definitely not 
correct and also the hydrostatic condition might be violated.

In Fig.~\ref{fig:sbd}, we show a map of the differences, $(I-M)/\sqrt{M}$,
between the unsmoothed exposure corrected image, $I$, and the two-dimensional
distribution of a smoothed model, $M$, that we choose to be a $\beta-$model
with parameters constrained over the radial range 0--400 kpc.
Note that positive excesses are present corresponding to (i) the central
20 arcsec--radius structure (i.e. cD galaxy plus filament) and (ii)
extended emission 50 arcsec from the centre towards North-East.
Negative values are predominant in the South-East direction.

\begin{figure}
\psfig{figure=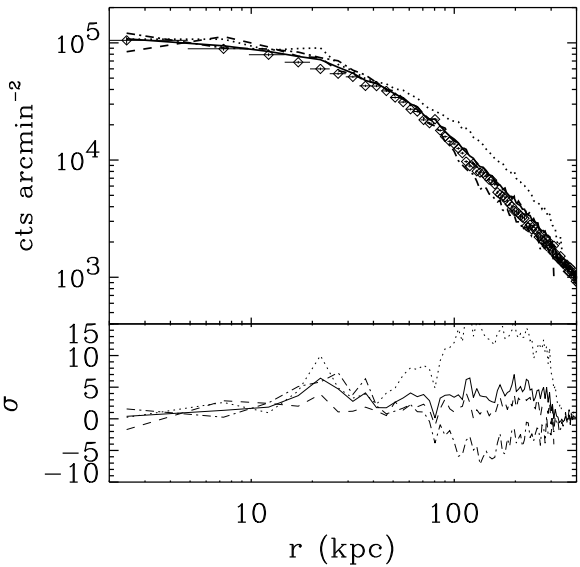,width=.5\textwidth}
\caption[]{
Brightness profiles extracted from four sectors ($x$ axis at 0\degr,
$y$ axis at 90\degr; cf. Fig.~\ref{fig:sbd}):
50\degr--135\degr (``Northern excess", dotted line),
135\degr--170\degr (dashed line), 170\degr--270\degr (``Southern deficit",
dot-dashed line), 270\degr--50\degr (diamonds).
The azimuthally averaged profile is represented with a solid line.
The residuals are plotted with respect to the ``270\degr--50\degr" profile.
} \label{fig:sb_sect}
\end{figure}

The evidence for regions with an excess/deficit in emission along the
North-South direction, but not aligned, becomes clear when
the brightness profiles extracted in these sectors are 
compared to the emission from westward of the centre
(Fig.~\ref{fig:sb_sect}).
Starting at about 50 kpc, the excess in the North-East and the deficit
in the South-East become evident with deviations respective to 
the West region of about $10 \sigma$.
This suggests that the core is in a non-relaxed status with indication
of merging activity. In Sect.~5.4, we discuss more on the 
dynamical status of the cluster core.

\subsection{X-ray colour profile}

\begin{figure}
\psfig{figure=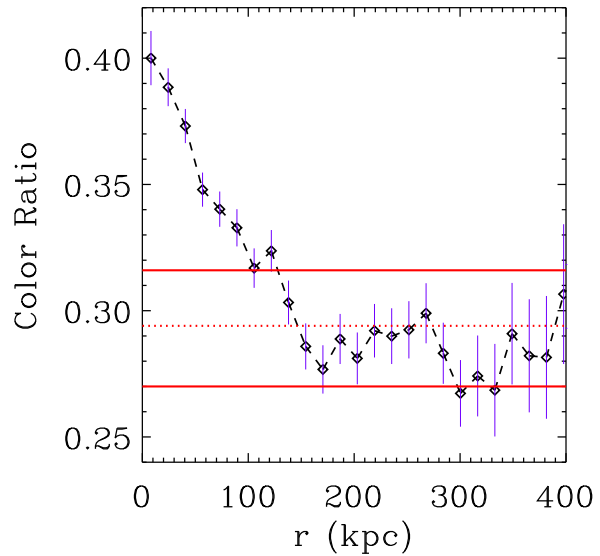,width=.5\textwidth}
\caption[]{Color Ratio [=(soft-hard)/(soft+hard)]
obtained using the X-ray surface brightness profiles
in the 0.5--1.5 keV and 1.5--7 keV band.
Note how these colour ratios flatten at about 70 arcsec,
equivalent to about 120 kpc.
} \label{fig:col_rat}
\end{figure}

To study the spatial distribution of the photons with respect 
to their energy,
we extract images in two X-ray `colours' (0.5--1.5 keV,
1.5--7 keV), correct them by the respective exposure maps
and subtract the background estimated from blank fields 
(see further comments in Sect.~5.2).
The surface brightness profiles are then obtained in bins
of 20 physical pixels (equal to 9.8 arcsec) and the color ratios 
determined (see Fig.~\ref{fig:col_rat})
 
The colour ratio rapidly decreases and flattens moving outwards,
indicating a hardening in the photon counts. We have quantified 
this in a non-parametric way, calculating a weighted average
(and relative error) of the colour ratio for all the points above
a given radius. Then, a distribution of the number of points 
enclosed within a given colour ratio $\pm 3 \sigma$ is obtained.
The values corresponding to (16, 50, 84) per cent of the 
cumulative function of this distribution is plotted
in Fig.~\ref{fig:col_rat}. The break between the ``most--probable''
value and the rapid increase of the profile in the centre 
is located at $r=72^{+40}_{-32}$ arcsec ($119^{+66}_{-53}$ kpc).

This excess of the relative amount of soft ($<$1.5 keV) photons
with respect to the harder counts population traces the region 
with a lower gas temperature.
Whether this gas is cooling or just following the behaviour of the 
gravitational potential is discussed in the following sections.

\section{Spectral analysis}

The spectra were extracted from the processed events-2 files, selecting 
in {\sc PI} space for energies between 0.5 and 7 keV to minimize the 
effects both of the uncertainties in the calibration of the observed soft X-rays and
of the cosmic rays that dominate the background components at energies
larger than 5 keV (see at the page {\tt http://asc.harvard.edu/
~cal/Links/~Acis/acis/~Cal\_prods/bkgrnd/~current/index.html}
maintained by M. Markevitch).
The background spectrum has been extracted from the same region 
of the CCD in exam from blank field exposures 
(namely {\tt aciss\_B\_s3\_bg\_evt\_060600.fits} and
{\tt aciss\_C\_s3\_bg\_evt\_191000.fits} for FP--110\degr 
and FP--120\degr, respectively)
adapted to our observations through the routine {\it make\_acisbg}
provided to the \chandra community by M. Markevitch.
As presented in Table~2, the total counts rate estimated in each
ring in the adopted 0.5--7 keV energy range is contaminated 
from the background by a percentage between less 
than 1 per cent (in the inner annulus) and 4 per cent 
(in the outer ring), making this correction lower than
statistical uncertainties. 


Redistribution Matrix Files (RMF) and Auxiliary Response Files (ARF)
were made by averaging over a $32\times32$ pixel-grid of calibration
files covering the back--illuminated chip 7, using the CIAO tools 
{\it mkrmf} and {\it mkarf} and using weighting factors equal to 
the number of counts in the source in the
region covered by the calibration.

In the present analysis, we use both single--phase and
multi--phase models, considering that we are mapping the very
central part of the cluster core. The single--phase model
just assumes an emission from an optically--thin plasma 
({\sc Mekal} --Kaastra 1992, Liedhal et al. 1995-- 
in XSPEC v.~11.1.0 --Arnaud 1996) absorbed
by a column density that we have left free to vary
(hereafter, model {\it abs1T}; the reference column density
is the Galactic value of $1.2 \times 10^{20}$ cm$^{-2}$ 
from radio HI maps in Dickey \& Lockman 1990).

The multi--phase model combines thermal emission from the 
ambient, outer gas with a continuous distribution of gas states 
represented by a cooling--flow model (Johnstone et al. 1992).
An absorbing column density fixed to the Galactic value
is put in front of the combined model.
The cooling--flow component is also intrinsically absorbed by 
uniformly distributed amount of hydrogen atoms per cm$^2$ 
at the cluster redshift that is left free to vary during
the search for a minimum $\chi^2$ ({\it absCF} model).
The temperature and metallicity of the cooling--flow model
are fixed to be equal to the ones of the thermal component. 
All the absorption of X-rays due to the interstellar medium has
been parametrized using the T\"ubingen-Boulder model ({\sc tbabs} 
in XSPEC v.~11.1.0; Wilms, Allen \& McCray 2000).

\begin{figure*}
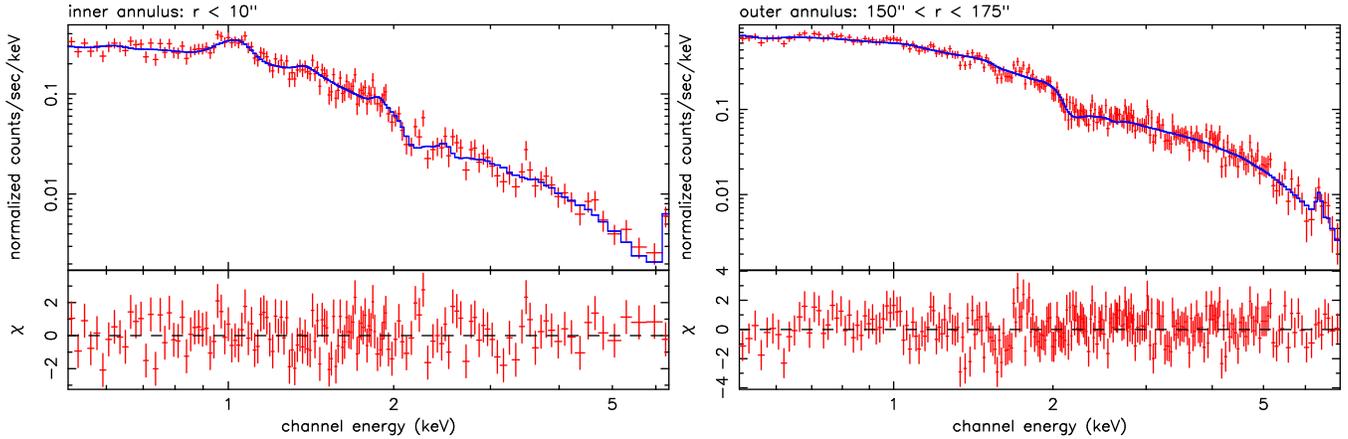

\hbox{
\psfig{figure=fig6a.ps,width=.5\textwidth,angle=-90}
\psfig{figure=fig6b.ps,width=.5\textwidth,angle=-90}
}
\caption[]{Observed spectra and best-fit absorbed {\sc Mekal} model
for two representative annular regions.
} \label{fig:spec}
\end{figure*}

\begin{table*}
\caption[]{Results of the spectral analysis in annuli.
In the second column, we quote the net count rate in the 0.5--7 keV 
energy range and, in parenthesis, the percentage of the total
counts that are associated to the source.
The models used are: {\it abs1T} = {\tt tbabs(mekal)} and {\it absCF} =
{\tt tbabs(mekal+ztbabs(cfmodel))}.
The {\it absCF} has the column density 
fixed to the Galactic value of $1.2 \times 10^{20}$ cm$^{-2}$, 
whereas {\it abs1T} considers $N_{\rm H}$ as free parameter.
$\dot{M}$ (in $M_{\odot}$ yr$^{-1}$) and $\Delta N_{\rm H}$ 
(in $10^{22}$ cm$^{-2}$) are integrated values from the 
circular region that includes all the inner annuli.
The comparison between the null-hypothesis probability 
of the {\it absCF} model versus {\it abs1T} is in Fig.~\ref{fig:com_spec}.
The metallicity refers to the solar photospheric values 
in Anders \& Grevesse (1989).
}
\begin{tabular}{r@{\hspace{.7em}} c@{\hspace{.7em}} c@{\hspace{.7em}}
c@{\hspace{.7em}} c@{\hspace{.7em}} c@{\hspace{.7em}} c@{\hspace{.7em}}
c@{\hspace{.7em}} c@{\hspace{.7em}} c@{\hspace{.7em}} c@{\hspace{.7em}} 
c@{\hspace{.7em}}}
Ann    & Net cts rate &  
   &  \multicolumn{3}{c}{ {\it abs1T} }  &
   & \multicolumn{5}{c}{ {\it absCF} }  \\ 
$"$ (kpc) & cts s$^{-1}$ (\%) & & $T$/keV & $Z/Z_{\odot}$ & $\chi^2$ (d.o.f.) &
    &  $T$/keV & $Z/Z_{\odot}$ & $\dot{M}$ & $\Delta N_{\rm H}$  &  
  $\chi^2$ (d.o.f.) \\ 
0--10 (0--17)  & 0.362$\pm$0.004 (99.9) & & 
$3.36^{+0.12}_{-0.12}$ & $0.63^{+0.09}_{-0.07}$ & 159.2 (141) && 
 $3.83^{+0.16}_{-0.33}$ & $0.75^{+0.12}_{-0.14}$ & $7.9^{+2.4}_{-2.5}$ & 
 $0.18^{+0.24}_{-0.05}$ & 156.3 (140) \\ 
10--25 (17--41)  & 1.254$\pm$0.008 (99.8) &&
$3.91^{+0.11}_{-0.10}$ & $0.56^{+0.05}_{-0.05}$ & 306.8 (241) && 
 $4.23^{+0.09}_{-0.10}$ & $0.60^{+0.05}_{-0.05}$ & $36.0^{+7.4}_{-7.9}$ & 
 $0.25^{+0.06}_{-0.05}$ & 300.7 (240) \\
25--40 (41--66)  & 1.506$\pm$0.009 (99.7) &&
$4.86^{+0.12}_{-0.10}$ & $0.71^{+0.06}_{-0.06}$ & 323.8 (275) && 
 $5.28^{+0.15}_{-0.14}$ & $0.75^{+0.06}_{-0.06}$ & $38.3^{+6.5}_{-6.3}$ & 
 $0.12^{+0.03}_{-0.03}$ & 312.2 (274) \\
40--55 (66--91)  & 1.337$\pm$0.008 (99.6) && 
$5.23^{+0.15}_{-0.23}$ & $0.43^{+0.06}_{-0.04}$ & 262.5 (262) && 
 $5.86^{+0.19}_{-0.17}$ & $0.45^{+0.07}_{-0.05}$ & $58.7^{+8.8}_{-7.7}$ & 
 $0.12^{+0.02}_{-0.03}$ & 248.9 (261) \\
55--75 (91--124)  & 1.474$\pm$0.009 (99.3) && 
$5.55^{+0.17}_{-0.13}$ & $0.38^{+0.06}_{-0.04}$ & 308.6 (281) && 
 $5.91^{+0.19}_{-0.15}$ & $0.40^{+0.06}_{-0.04}$ & $73.9^{+9.5}_{-8.5}$ & 
 $0.11^{+0.02}_{-0.02}$ & 305.6 (280) \\
75--100 (124--165) & 1.509$\pm$0.009 (98.8) && 
$5.94^{+0.15}_{-0.17}$ & $0.44^{+0.05}_{-0.05}$ & 369.4 (290) && 
 $6.33^{+0.15}_{-0.29}$ & $0.46^{+0.05}_{-0.06}$ & $90.3^{+9.1}_{-10.5}$ & 
 $0.11^{+0.02}_{-0.02}$ & 367.4 (289) \\
100--125 (165--206) & 1.255$\pm$0.008 (98.1) && 
$6.10^{+0.21}_{-0.12}$ & $0.37^{+0.06}_{-0.06}$ & 294.2 (274) && 
 $6.84^{+0.39}_{-0.37}$ & $0.38^{+0.06}_{-0.06}$ & $100.8^{+13.4}_{-11.8}$ & 
 $0.12^{+0.02}_{-0.02}$ & 290.6 (273) \\
125--150 (206--248) & 1.062$\pm$0.008 (97.4) && 
$6.34^{+0.25}_{-0.20}$ & $0.35^{+0.07}_{-0.06}$ & 286.8 (258) && 
 $7.43^{+0.55}_{-0.45}$ & $0.39^{+0.08}_{-0.08}$ & $115.6^{+12.1}_{-9.5}$ & 
 $0.11^{+0.01}_{-0.01}$ & 279.4 (257) \\
150--175 (248--289) & 0.887$\pm$0.007 (96.3) && 
$6.49^{+0.43}_{-0.39}$ & $0.18^{+0.07}_{-0.07}$ & 265.1 (239) && 
 $8.97^{+0.87}_{-0.94}$ & $0.23^{+0.08}_{-0.06}$ & $132.9^{+11.6}_{-11.6}$ & 
 $0.10^{+0.01}_{-0.01}$ & 250.3 (238) \\
\end{tabular}
\end{table*}

\begin{figure*}
\hbox{
\psfig{figure=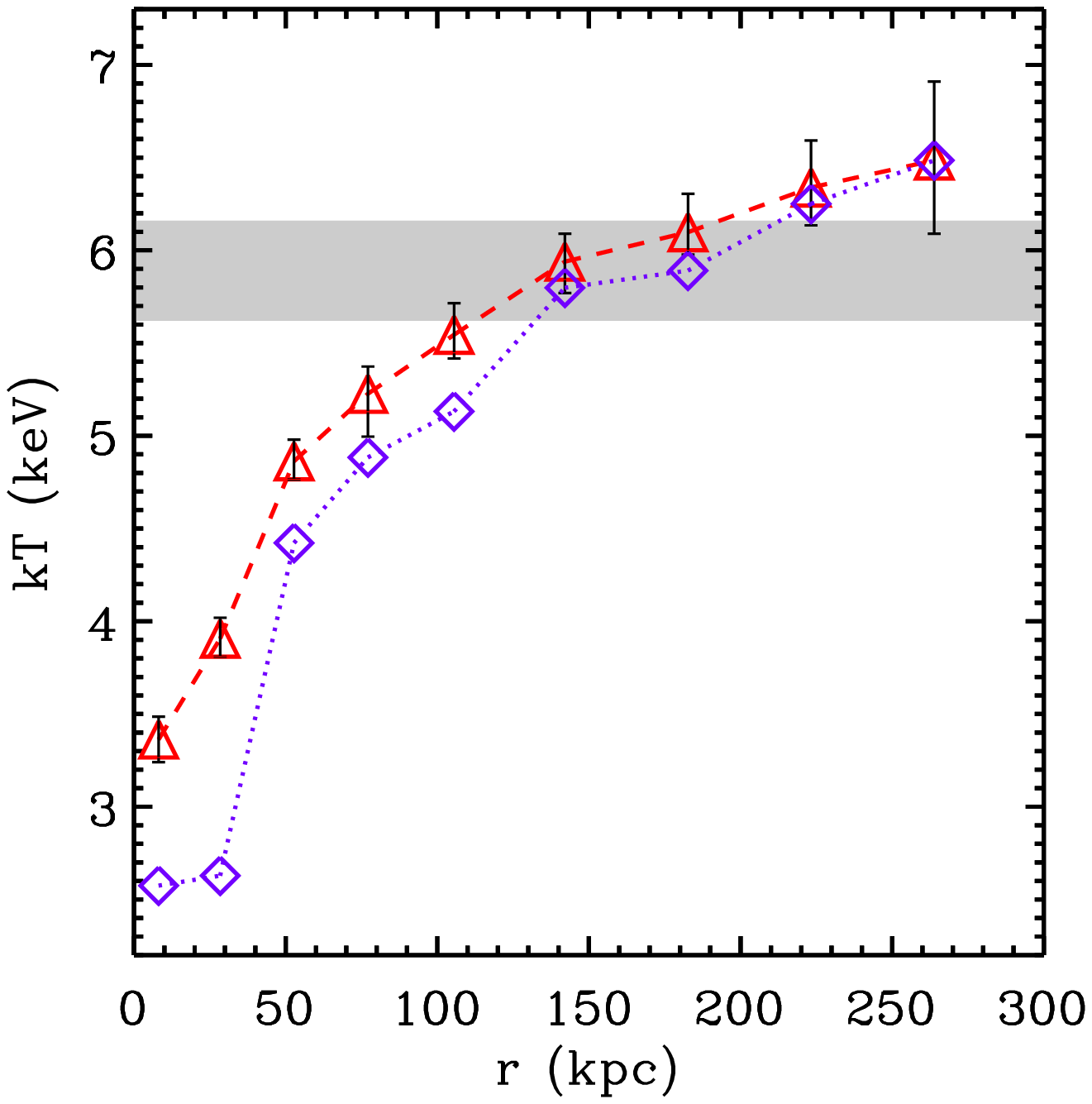,width=.5\textwidth}
\psfig{figure=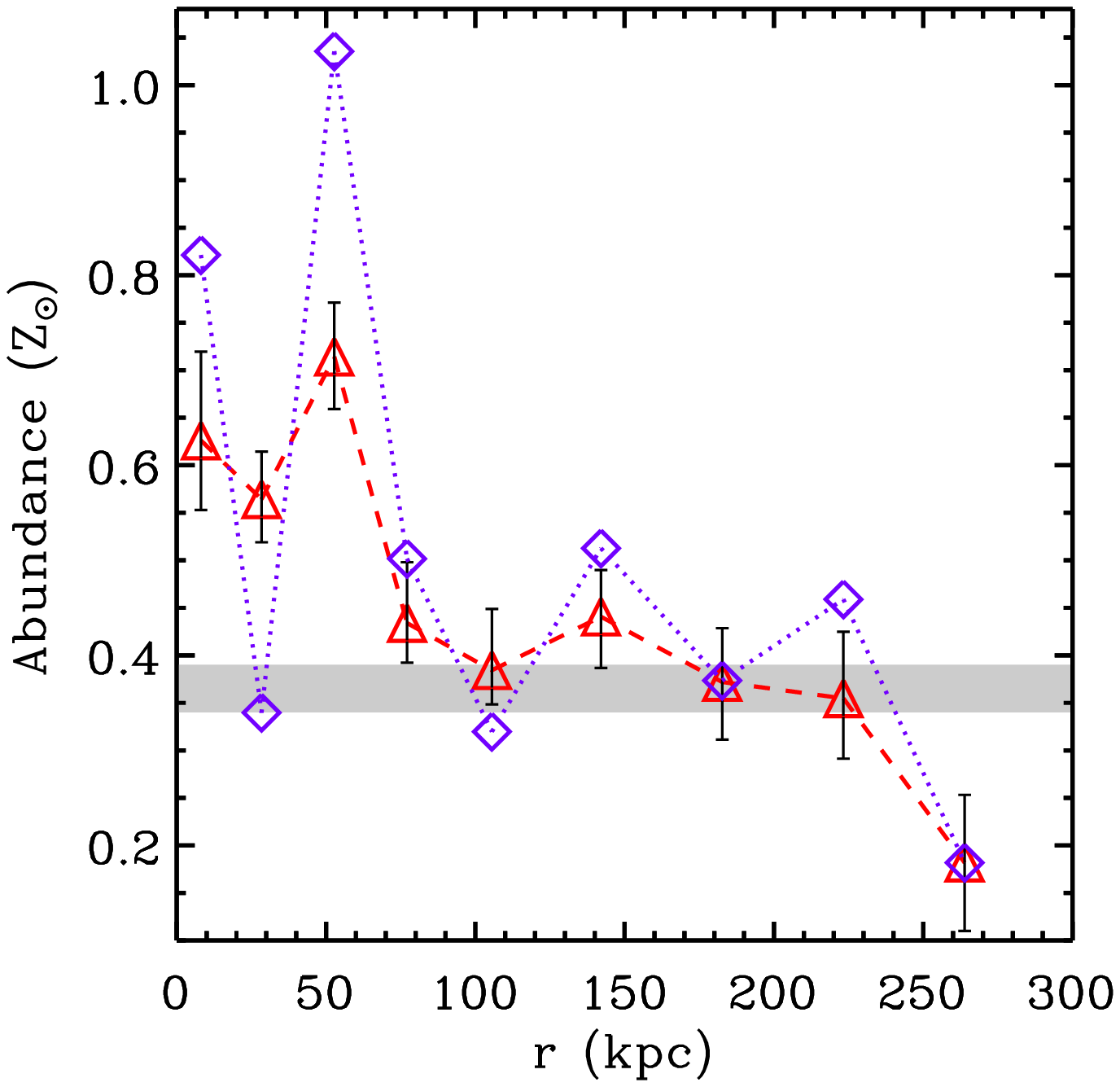,width=.5\textwidth}
} \vspace*{-.7cm} \hbox{
\psfig{figure=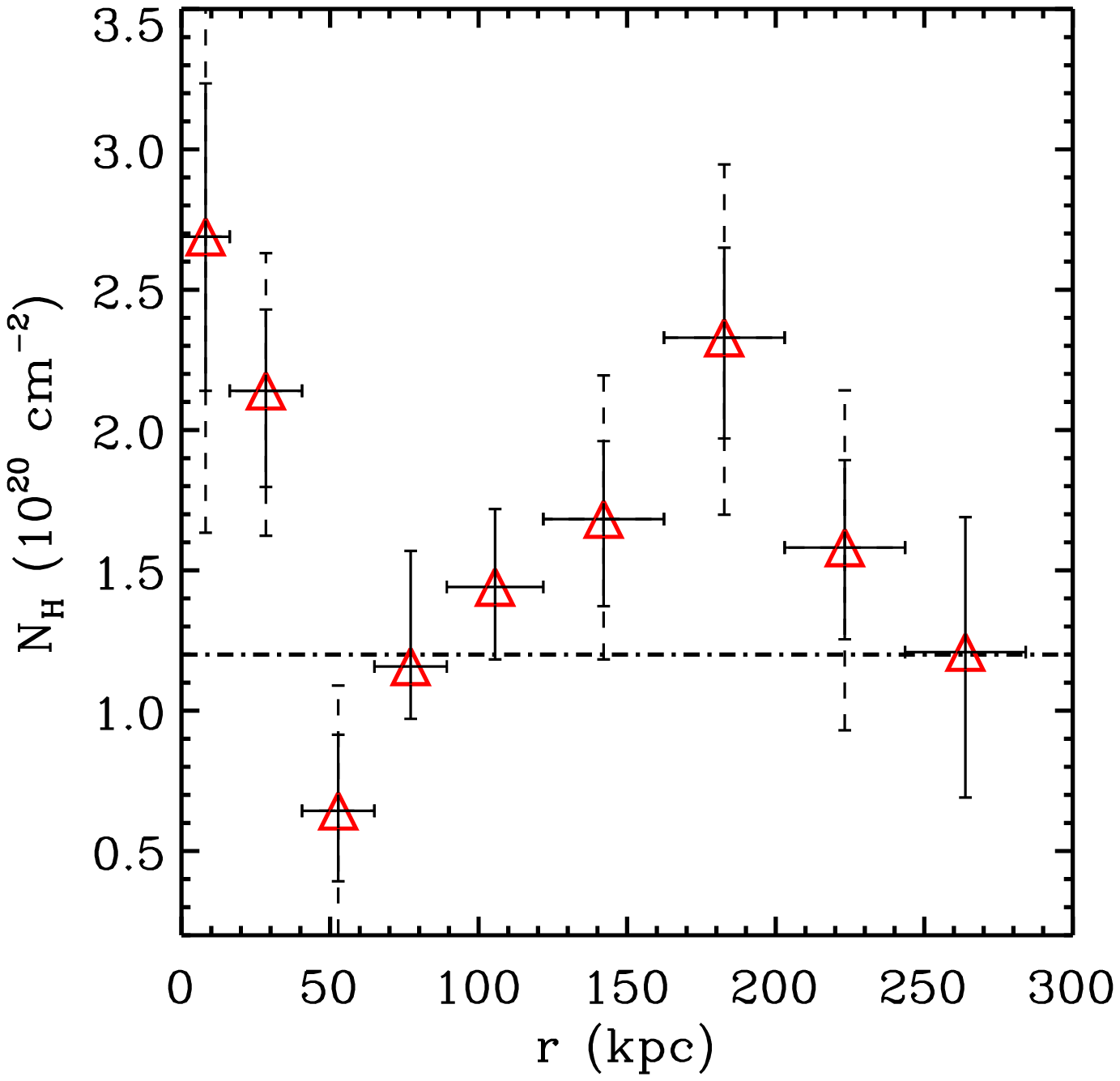,width=.5\textwidth}
\psfig{figure=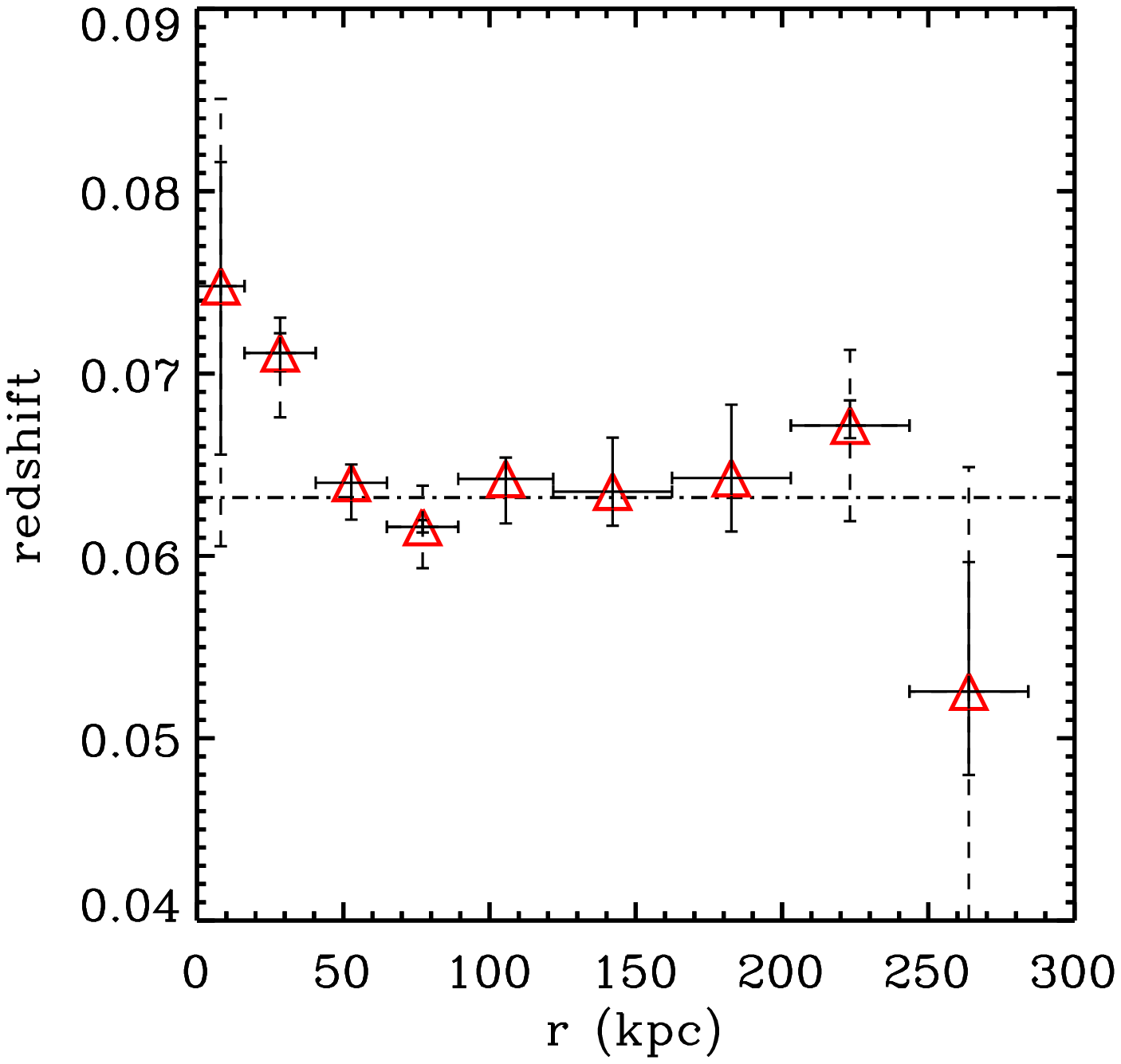,width=.5\textwidth}
}
\caption[]{Best-fit spectral results applying 
a single absorbed {\sc Mekal} model ({\it triangles}; solar
metallicity from Anders \& Grevesse 1989)
and the corresponding deprojected results ({\it diamonds}).
The error bars are at $1 \sigma$ level.
The shaded regions represent the uncertainties at the 90 per cent
level of confidence from the \asca analysis in Allen (2000).
The dashed error bars refer to the 90 per cent
confidence level ($\Delta \chi^2$ = 2.71 on the interesting parameter). 
The dot--dashed lines indicate the Galactic absorption ($N_{\rm H} =
1.2 \times 10^{20}$ cm$^{-2}$, Dickey \& Lockman 1990) and
the optical determination of the redshift at 0.0632 (Girardi et al. 1998),
respectively.
} \label{fig:res_spec}
\end{figure*}

Two representative spectra are shown in Figure~\ref{fig:spec} with 
the respective {\it abs1T} model.
In Table~2, we quote the best-fit results for the gas temperature and
metallicity obtained with a single-phase plasma model ({\it abs1T})
and with a multi-phase component ({\it absCF}).
The minimun $\chi^2$ and the number of degrees
of freedom for these models are quoted in Table~3.
In Figure~\ref{fig:res_spec}, we plot the best-fit values 
obtained with the model {\it abs1T} for 
some interesting quantities and, where available, the 
region of uncertainty at the 90 per cent confidence level from 
the analysis of \asca data in Allen (2000).
Steep temperature and metallicity profiles are present within the 
inner 300 kpc. The emission-weighted 
gas temperature increases from 3.4 to 6.5 keV, whereas the metal
abundance decreases with radius from about 0.7 times the solar
photospheric values in Anders \& Grevesse (1989) to $\sim$0.2.
The column density is consistent with the Galactic value above
50 kpc, but it is higher by a factor of 2 in the inner two annuli.
A remarkable good agreement is present between the redshift measured 
independently in each ring and the optical estimate from Girardi et al. 
(1998).  

\begin{figure*}
\hbox{
\psfig{figure=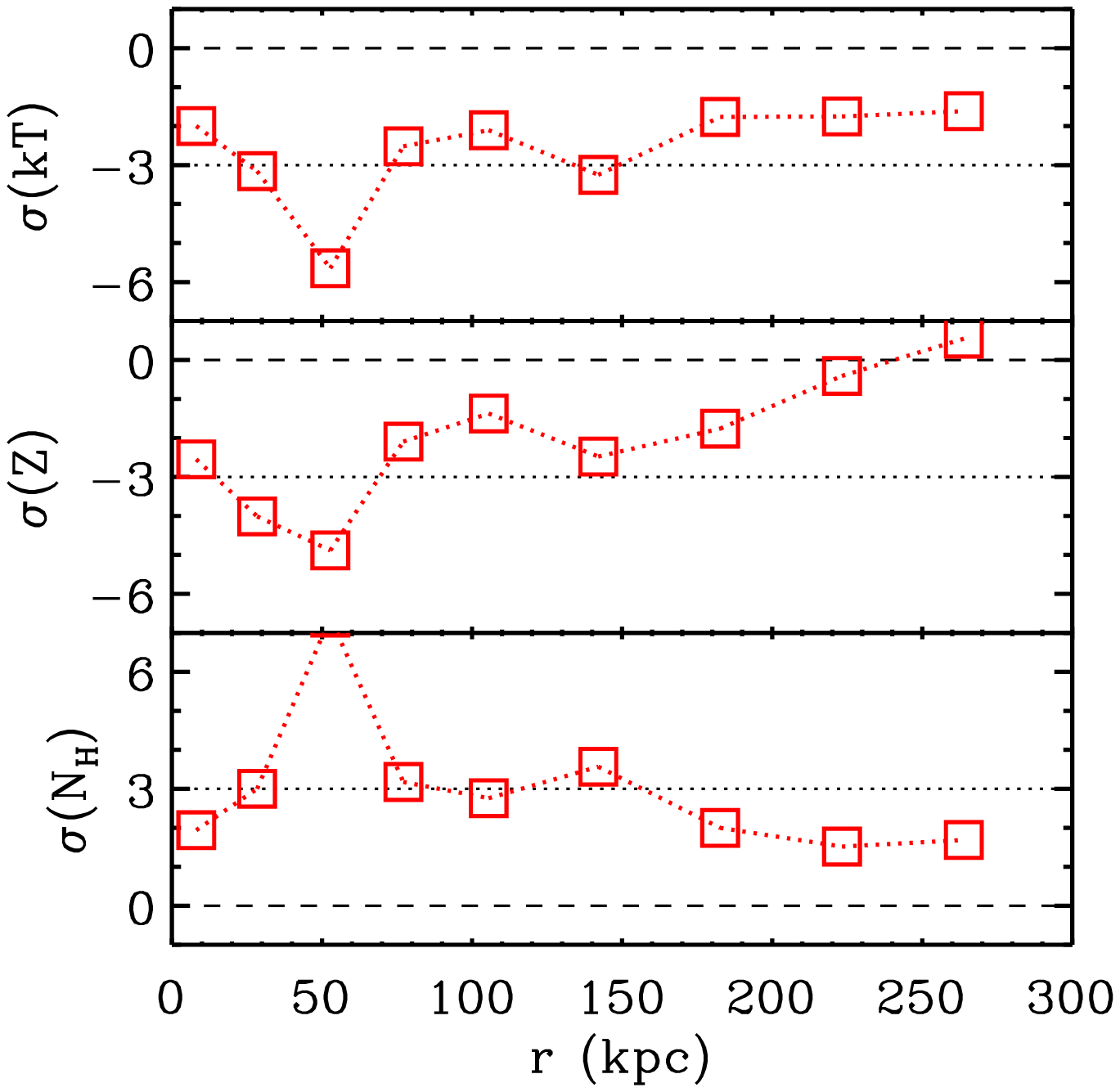,width=.5\textwidth}
\psfig{figure=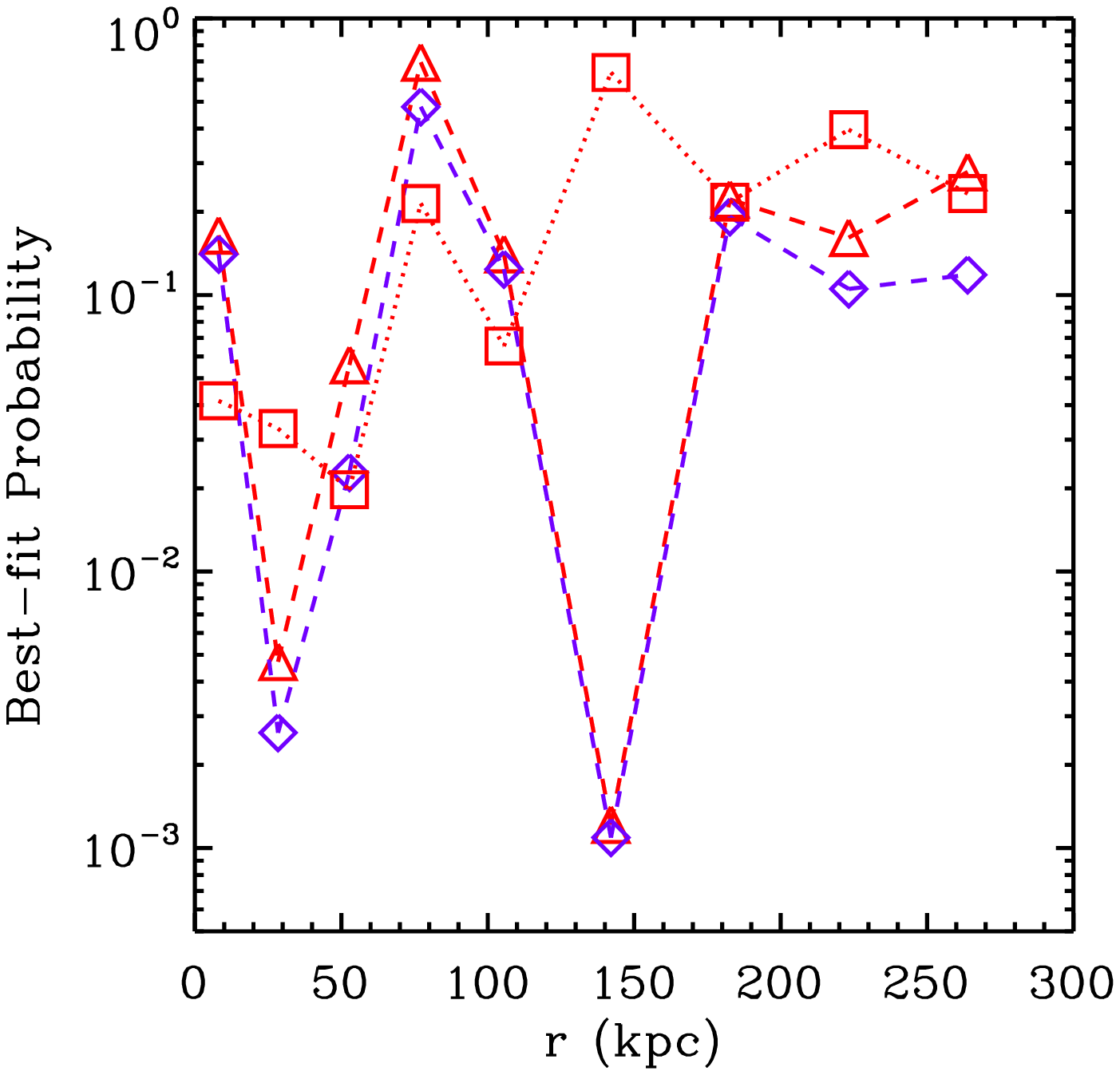,width=.5\textwidth}
}
\caption[]{
{\bf Left panel} This panel shows the discrepancy (in $\sigma$) 
between the best-fit results using the {\it abs1T} model
between the datasets with Focal Plane temperature of --110\degr
(calibration files of the December 1999 release) and 
FP--120\degr calibration files of the August 2001) currently
used and plotted in Fig.~\ref{fig:res_spec}:
$(data_{-110} - data_{-120})/\sqrt{\sigma_{-110}^2+\sigma_{-120}^2}$.
{\bf Right panel} The null hypothesis probability of the 
best-fit results is shown for each annulus considered for the 
following {\it dataset/model}: 
``--120\degr/{\it absCF}'' ({\it triangles}),
``--120\degr/{\it abs1T}'' ({\it diamonds}), 
``--110\degr/{\it abs1T}'' ({\it squares}).
} \label{fig:com_spec}
\end{figure*}

In Fig.~\ref{fig:com_spec} and Table~3, we show that (i)
the null-hypothesis probability on the goodness of the fits 
is larger than 0.1 on 6 out of nine annuli for both the {\it abs1T}
and {\it absCF} models with the worst reduced $\chi^2$ of about 1.27
in the second and sixth annulus, and (ii) that 
the cooling flow model, {\it absCF}, is equivalent to the 
single-phase model, {\it abs1T}, from a statistical point of view
when the F (Fisher) test is considered 
(e.g. Bevington \& Robinson 1992; on how a single-phase gas
can mimic a multi-phase medium through projection effects
see, e.g., Ettori 2001).

In the same Fig.~\ref{fig:com_spec}, we compare also
the spectral results for the 
FP--110\degr and FP--120\degr datasets. 
Gas temperatures, metallicities and column 
densities show differences generally between 1 and 3 $\sigma$
and a clear trend with temperature and metal abundance estimates
higher and column density values lower in the
FP--120\degr dataset.

%
%

\begin{table*}
\caption[]{Values of the minimum $\chi^2$ 
(degrees of freedom in parenthesis) for each spectrum 
investigated in the present work for the following models:
{\it abs1T} = {\tt tbabs(mekal)}; {\it abs1TZ} = {\tt tbabs(vmekal)};
{\it absCF} = {\tt tbabs(mekal+ztbabs(cfmodel))}.
The column density in {\it absCF} is 
fixed to the Galactic value of $1.2 \times 10^{20}$ cm$^{-2}$, 
whereas the other models consider $N_{\rm H}$ as free parameter.
The ratio $F$ between the reduced $\chi^2$ obtained from {\it abs1T} and
{\it absCF} is distributed as an F-distribution with $m$ and $n$ 
degrees of freedom, where $m$ and $n$ are equal to the d.o.f.
in {\it abs1T} and {\it absCF}, respectively 
(Bevington \& Robinson 1992, pag.~205).
The probability $P$ that a random variable is larger than $F$
is quoted with the required value, $F_{95}$, 
of $F$ to obtain a level of significance
of the 95 per cent.
The null-hypothesis probabilities of the best-fit results 
obtained with {\it absCF}, {\it abs1T} and {\it abs1TZ} 
are shown in Fig.~\ref{fig:com_spec} and Fig.~\ref{fig:metals}.
}
\begin{tabular}{rccccccc}
Ann  &  & {\it abs1T} & {\it abs1TZ} & {\it absCF} &  
  \multicolumn{3}{c}{F test} \\
 $"$ (kpc) &  & \multicolumn{3}{c}{ $\chi^2$ (d.o.f.) } & $F$ & $P$ & $F_{95}$\\
0--10 (0--17)  & &  
 159.2 (141) & 153.1 (137) & 156.3 (140) & 1.01 & 0.47 & 1.32 \\ 
10--25 (17--41)  &  & 
 306.8 (241) & 300.1 (237) & 300.7 (240) & 1.02 & 0.45 & 1.24 \\ 
25--40 (41--66)  &  &
 323.8 (275) & 318.2 (272) & 312.2 (274) & 1.03 & 0.39 & 1.22 \\ 
40--55 (66--91)  &  & 
 262.5 (262) & 249.2 (258) & 248.9 (261) & 1.05 & 0.34 & 1.23 \\
55--75 (91--124)  &  & 
 308.6 (281) & 306.3 (277) & 305.6 (280) & 1.01 & 0.48 & 1.22 \\
75--100 (124--165) &  & 
 369.4 (290) & 354.8 (286) & 367.4 (289) & 1.00 & 0.49 & 1.21 \\ 
100--125 (165--206) &  & 
 294.2 (274) & 285.8 (270) & 290.6 (273) & 1.01 & 0.47 & 1.22 \\  
125--150 (206--248) &  & 
 286.8 (258) & 279.0 (254) & 279.4 (257) & 1.02 & 0.43 & 1.23 \\
150--175 (248--289) &  & 
 265.1 (239) & 252.4 (235) & 250.3 (238) & 1.05 & 0.34 & 1.24 \\  
\end{tabular}
\end{table*}

\section{Deprojection analysis of the spectral and imaging data}

In this Section, we discuss the constraints on the physical quantities 
obtained from the deprojection analysis applied to both the best-fit 
results of the spectral analysis presented in the previous section
(hereafter {\it spectral deprojection}) and the surface brightness profile
({\it spatial deprojection}).

The deprojection analysis makes the assumption that the emission is 
spherically symmetric and the X-ray emitting plasma is in hydrostatic 
equilibrium with the underlying gravitational potential, 
which is likely to be reasonable on intermediate to larger scales
considering that the sound crossing time in the inner 100 kpc is less
than $10^8$ years. 

Considering that we are not able to distinguish between
single phase and cooling flow models in the fitting analysis
of our spectra as discussed in Section~4, we adopt hereafter
for simplicity the single phase description to infer 
the properties of the intracluster medium.

\subsection{Spectral deprojection}
The physical quantities constrained from the projected spectra
need to be translated to their 3-dimensional values to allow us 
to investigate the proper characteristics of the X-ray emitting
plasma. Fitting a thermal model to a spectrum obtained 
collecting X-ray counts in rings provides, for each annulus,
(i) an estimate for the Emission Integral, $EI = \int n_{\rm e}
n_{\rm p} dV = 0.82 \int n_{\rm e}^2 dV$, 
through the normalisation $K$ of the model, $K = \frac{10^{-14}}
{4 \pi d_{\rm ang}^2 (1+z)^2} EI$ (see {\sc Mekal} model in XSPEC); 
(ii) a direct measurement of the emission-weighted gas temperature,
$T_{\rm ring}$, metal abundance, $Z_{\rm ring}$, and 
luminosity, $L_{\rm ring}$.
The purpose of the deprojection is, for example, to recover
the value of the gas temperature in shells, 
$T_{\rm shell} \equiv T_i$, that is defined as
\begin{equation}
T_{\rm ring} \equiv T_j = \frac{ \sum_{i, {\rm outer \ shell}}^{i=j}
T_i w_{ij} }{ \sum_{i, {\rm outer \ shell}}^{i=j} w_{ij} }
\label{eq:tem}
\end{equation}
where $w_{ij} = L_i \times Vol(i,j)/Vol(i) = \epsilon_i Vol(i,j)$ 
provides the luminosity
for a given shell $i$ with volume $Vol(i)$ weighted by the part 
of this volume projected on the ring $j$, $Vol(i,j)$.
Using this notation, it is simple to note that 
$L_{\rm ring} \equiv L_j = \sum_{i, {\rm outer \ shell}}^{i=j} 
\epsilon_i Vol(i,j) = \sum_{i, {\rm outer \ shell}}^{i=j} w_{ij}$. 

From Kriss, Cioffi \& Canizares (1983; see also McLaughlin 1999),
the volume of each shell observed through each ring adopted in the spectral
analysis can be evaluated and a matrix, ${\bf Vol}$, can be built
with components equal to the parts the volume of the shells (rows $i$)
seen at each ring (or annuli; column $j$).

The deprojected physical quantities can be then obtained through 
the following matrix products (shown by the symbol $\#$):
\begin{equation}
\begin{array}{l}
n_{\rm e} = \left[ ({\bf Vol}^T)^{-1} \# (EI/0.82) \right]^{1/2} \\
\epsilon = ({\bf Vol}^T)^{-1} \# L_{\rm ring} \\
\epsilon T_{\rm shell} = ({\bf Vol}^T)^{-1} \# (L_{\rm ring} T_{\rm ring}) \\
\epsilon Z_{\rm shell} = ({\bf Vol}^T)^{-1} \# (L_{\rm ring} Z_{\rm ring}), 
\end{array}
\end{equation}
where $({\bf Vol}^T)^{-1}$ indicates that the matrix is firstly
transposed and then inverted.
The emission due to the shells projected along the line of sight 
but with the corresponding annuli outside the field-of-view is taken 
into account with an edge correction factor estimated assuming a
power law distribution of the emission proportional to $r^{-4}$
(cfr. equation A8 in the appendix of McLaughlin 1999).

We have applied this technique to the single-phase results of our
spectral analysis (model {\it abs1T} in Sect.~4). 
In rings where part of the flux was masked for the presence of 
a point-source, we correct the normalisation $K$ by the relative
amount of area not considered.
(We note that, if we fit simultaneously the nine spectra
with a combination of single phase models properly weighted with 
the correspondent volume of the shell observed at each ring, 
we measure a $\chi^2$ of 2604.3 with 2278 degrees of freedom
that is statistically undistinguishable from the sums of the
results in Table~2 of $\chi^2=$2576.4 and 2261 degrees of freedom).

The deprojected temperature gradient can then be compared 
with the output of the spatial deprojection code to constrain 
the gravitational potential and the mass deposition
rate.
The error bars come from 100 Monte-Carlo simulations obtained 
from scattering the original projected input with respect to their
Gaussian error.

The deprojected gas temperature and metallicity 
are slightly steeper than the observed, projected profiles
(cf. Fig.~\ref{fig:res_spec}).
The gas temperature rises from $2.6^{+0.3}_{-0.5}$ to
$6.5^{+0.4}_{-0.3}$ keV at 270 kpc. The profile can 
be fitted ($\chi^2$=10.5, 7 d.o.f., P=0.16) with a power law:
$T_{\rm gas} = 2.74^{+0.41}_{-0.23} \times (r/{\rm 10 kpc})^{0.27
(-0.06, +0.04)}$ keV.
The metal (iron) abundance decreases from a central value of
about 0.8 times the solar abundance to $0.18 (\pm 0.07)$.

\subsection{Spatial deprojection}

\begin{figure*}
\hbox{
\psfig{figure=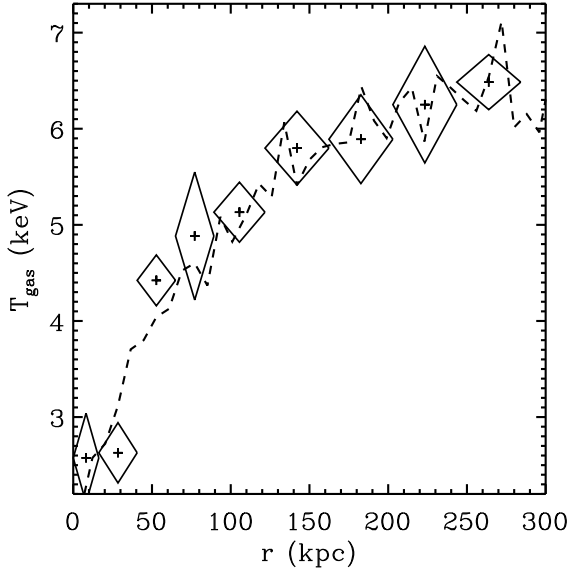,width=.5\textwidth}
\psfig{figure=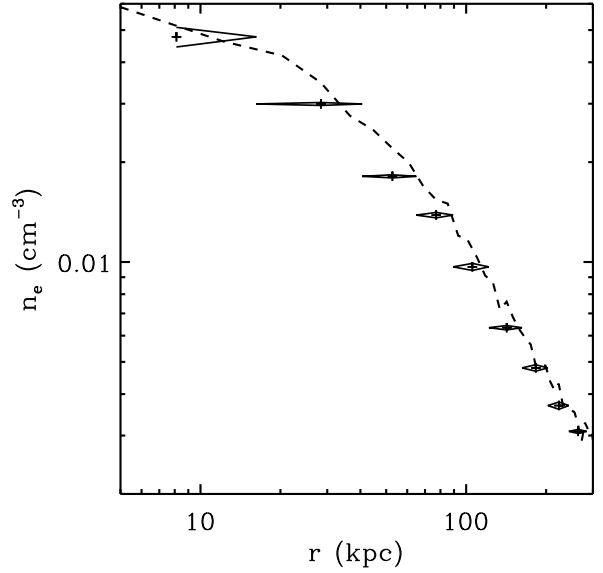,width=.5\textwidth}
}
\caption[]{Comparison between the {\it spectral} (diamonds with error
bars) and {\it spatial} (dashed line) deprojection
results on the gas temperature (left panel) and electron density
(right panel).
} \label{fig:nele_t}
\end{figure*}

To reach more firm conclusions on the state of the intracluster
plasma in the inner region, we have combined the results
of the spectral deprojection with the {\it spatial deprojection}
of the X-ray surface brightness profile, which has been 
done applying the code developed from Fabian et al. 
(1981; see details on the code in White, Jones \& Forman 1997).
The surface brightness profile has been obtained in the 0.5--7 
keV energy range with a bin-size of 9.8 arcsec (i.e. 20 times the 
original ACIS bin-size).
Considering that the cluster emission covers the whole CCD, 
we have estimated a background count-rate of about 1.5 $\times
10^{-6}$ cts s$^{-1}$ arcsec$^{-2}$ from the blank-field
used in the spectral analysis (Sect.~4).
We detect signal up to 4.4 arcmin ($\sim$ 440 kpc) from the assigned 
centre with $\ga 5 \sigma$ confidence. 

The results from the spectral deprojection analysis are used to constrain
the process of the spatial deprojection in the way described below
(and similar to the procedure adopted in Allen, Ettori \& Fabian 2001, 
Schmidt, Allen \& Fabian 2001).
In the spatial deprojection, there are 5 free parameters: (i)
the column density, $N_{\rm H}$, fixed to the averaged value observed 
in the spectral analysis of about $1.6 \times 10^{20}$ cm$^{-2}$;
(ii) the metallicity fixed to 0.3 times the solar value in 
accordance with the mean behaviour of the deprojected abundance 
profile; (iii) the outer pressure, fixed to the value
of $27.6 \times 10^4$ K cm$^{-3}$, extrapolated from the pressure 
profile estimated from the deprojected spectral 
temperature and gas density profiles;
(iv-v) the scale (core) radius and the velocity dispersion 
describing the assumed potential law [in the present case 
either an isothermal sphere (Binney \& Tremaine 1987, eqn.~4-124b) 
or a Navarro, Frenk \& White profile (NFW, 1997)].
The last two parameters are obtained from a $\chi^2$ minimization
between the gas temperature profile obtained from spectral
deprojection of the single-phase analysis results (cf. Sect.~4) 
and the output of the spatial deprojection code properly rebinned 
(cf. eqn.\ref{eq:tem}) to match the number of bins of the 
spectral profile.
(In the spatial deprojection, the errors are obtained from the 
distribution of the results of the deprojection of 100 Monte-Carlo 
simulations of the original surface brightness profile.)

The best-fit between the two temperature profiles are obtained 
for the following values of the two free parameters 
of the potential laws:
$r_{\rm c} = 0.10^{+0.01}_{-0.01}$ Mpc, 
$\sigma_{\rm IS} = \sqrt{(4/9) \pi G \rho_0 r_{\rm c}^2} =
658^{+14}_{-11}$ km s$^{-1}$ for the isothermal sphere;
$r_{\rm s}= 0.49^{+0.11}_{-0.13}$ Mpc, 
$\sigma_{\rm NFW} = \sqrt{(G M_{200})/(2 r_{200})} 
= \sqrt{50} \ H_0 \ r_{200} =
837^{+69}_{-85}$ km s$^{-1}$ when a NFW potential law is assumed.
Note that $\rho_0$ and $r_{\rm c}$ define the central matter density 
and the core radius, respectively, in the isothermal sphere, whereas
$M_{200}$ indicates the total mass within the radius $r_{200}$ where
an overdensity of 200 with respect to the background is reached.
Note also that the two velocity dispersions are related by 
the equation
$2 \sigma_{\rm NFW}^2 = (- d\ln \rho / d\ln r) \ \sigma_{\rm IS}^2$
(cfr. eqn.4-127b in Binney \& Tremaine 1987) 
and tend to be equal when regions extending at $r \ga 10 r_{\rm c}$
ca be considered (see Figure 4-8 in Binney \& Tremaine 1987;
in our case, we are mapping regions within $\sim 4 r_{\rm c}$).
However, we observe a lower $\chi^2$ for a potential with a flat
profile in the core, namely $\chi^2=$ 7.1 (with 7 degrees of freedom, P=0.41)
for the isothermal sphere against 16.5 (P=0.02) for a NFW profile.

In Fig.~\ref{fig:nele_t}, we show the gas temperature and density
profiles from the spectral deprojection analysis 
compared to those obtained from the best-fit results in the
spatial deprojection.
It is worth noting that the good agreement between the two gas
density profiles is due to the dominance of the density in the 
X-ray emissivity that allows an almost temperature-independent
deprojection, whereas we have required that the two temperature
profiles match, searching for a minimun $\chi^2$ in comparing
these profiles with the parameters of the gravitational potential
left free to vary.
Moreover, this comparison shows the self-consistency of our results
and provides us with a cooling time and a mass deposition rate
profiles with resolution of the imaging data 
(see discussion in Section~5.5).

\begin{figure*}
\psfig{figure=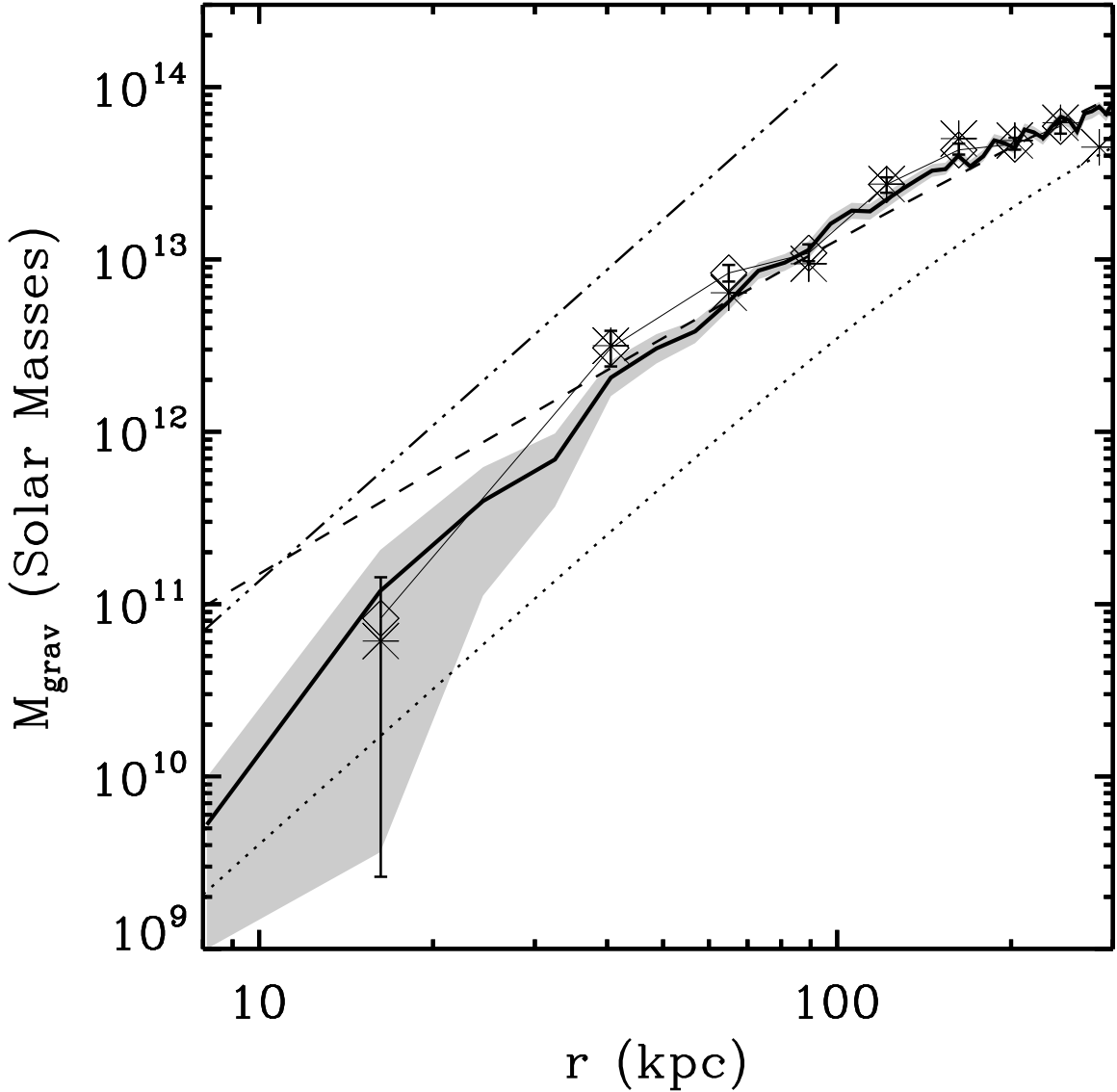,width=\textwidth}
\caption[]{Gravitational mass profiles.
The solid line is the mass profile obtained through
the hydrostatic equilibrium equation applied to the results
from the spatial deprojection on the gas temperature and density
profiles (see Sect.~5.2 for details).
The {\it diamonds} are the total mass values obtained
using the deprojected spectral temperature and density measures.
The {\it asterisks} are the total mass values obtained
from the spectral deprojection when the X-ray center is fixed
on the cD galaxy.
These values are compared with the best-fit results on \pspc data
(Ettori \& Fabian 1999) using a $\beta-$model (dotted line) and a
gas NFW model (dashed line).
The three-dots-dash line indicates the upper limit from an assumed
tidal shear in the $H_{\alpha}$ filament due to the central
cluster potential.
} \label{fig:dm}
\end{figure*}

\subsection{Cluster mass profile in the core}

In our spectral analysis, we are able to resolve in 5 radial bins 
the gas temperature profile in the inner 100 kpc radius with a relative
uncertainty less than 5 (10) per cent for the projected (deprojected)
temperature. 
This enables us to investigate the mass profile in the cluster core
assuming that the intracluster medium is there in hydrostatic equilibrium
with the dark matter potential. 

We make use of the gas temperature and electron 
density values estimated from the deprojected best-fit spectral results
obtained with a single-phase modeli (see Section~5.1). 
Then, we apply the equation of hydrostatic 
equilibrium between the gravitational potential and the intracluster
plasma to estimate the mass profile, $M_{\rm X}$,
\begin{equation}
M_{\rm X} = -\frac{T r^2}{G \mu m_{\rm p}} \left( \frac{d\ln T + d \ln n} 
{dr} \right).
\label{eq:mx}
\end{equation}
In doing this, we interpolate the gas temperature profile 
with a polynomial of 2nd order to smooth over any fluctuations 
for the results from spectral deprojection.
Moreover, we estimate the mass profile using a smoothed 
temperature and density profiles obtained from the best-fit 
results of the spatial deprojection (see Section~5.2).
We assume a relative error propagated from the results
of the spectral deprojection of about 15 per cent consistent with
1.5 times the uncertainty on the gas temperature measurements.
In Fig.~\ref{fig:dm}, we compare these dark matter profiles 
with the best-fit results obtained fitting the outer region of
the cluster emission observed with \pspc (for details see 
Ettori \& Fabian 1999) both with a $\beta-$model 
(Cavaliere \& Fusco-Femiano 1976; $r_{\rm c}$ =0.25 Mpc, $\beta$ = 0.75)
and a gas model obtained from the NFW potential law
($r_{\rm s}$ =0.77 Mpc, $\eta$ = 10.70).

\begin{figure}
\psfig{figure=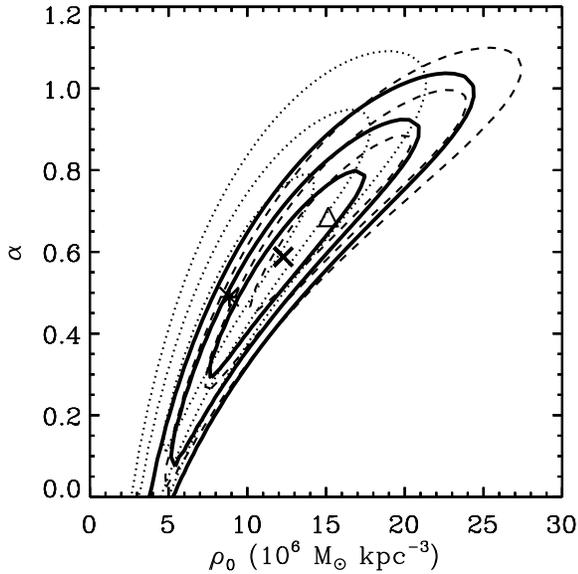,width=.5\textwidth}
\caption[]{Contours of the probability distribution
($\Delta \chi^2$= 2.3, 6.2, 11.8 that correspond to 1, 2, 3
$\sigma$ for two interesting parameters)
using the power-law model in eqn.~\ref{eq:mgrav} on the mass
profile from spectral deprojection results ({\it diamonds} in Fig.
\ref{fig:dm}) between 10 and 100 kpc ({\it solid line} and {\it cross}).
The {\it dotted line} ({\it asterisk}) is for the mass profile
from X-ray analysis centered on the cD galaxy.
The {\it dashed lines} ({\it triangle})
indicate the confidence levels when the mass
is increased by 40 per cent in the inner two bins only.
} \label{fig:dm_pow}
\end{figure}

To assess the shape of the underlying gravitational potential 
in the central region, we describe the dark matter profile
with a power law expression, $\rho_{\rm grav} = \rho_0 (r/r_0)^{-\alpha}$,
integrate it over the volume
\begin{equation}
M_{\rm grav} = \int_0^R 4\pi \rho_{\rm grav} r^2 dr =
\frac{4\pi \rho_0 r_0^3}{3-\alpha} \left( \frac{R}{r_0} \right)^{3-\alpha}
\label{eq:mgrav}
\end{equation}
and fit this mass profile to the one from the spectral deprojection
results (see Fig.~\ref{fig:dm}) between 10 and 100 kpc, where
the discrepancy among the different forms of the potential 
is more significant.
We measure $\alpha = 0.59^{+0.12}_{-0.17}$ (in the range 0.27--0.81 at the
90 per cent confidence level; Fig.~\ref{fig:dm_pow}),
that indicates a remarkable flat profile suggesting the presence of a core.
In particular, a value of 1 (typical for cuspy dark matter profile like
NFW) is excluded at more than 3 $\sigma$. 
In other words, we expect four times more than the observed mass
in the inner bin, if we assume a power law index of 1 and a fixed mass
at 100 kpc.
On the other hand, the deviation from a flat core (i.e. $\alpha =$ 0)
makes the mass profile lie apart from that expected given a King profile.
The presence of this ``shoulder'' has been detected also by
Xu et al. (1998) from \asca and \pspc data and
is now resolved with \chandra.
Finally, from the presence of a shear in the $H_{\alpha}$ filament
southward the cD galaxy, Hu, Cowie and Wang (1985) conclude
that the central gravitational potential is not too deep.
Following their arguments on the tidal shearing of the filament
in the cluster potential, we obtain an upper limit on the
central mass (assuming constant density) consistent with our
results and that excludes any dark matter density profile
sharper than the NFW form below 10 kpc (see Fig.~\ref{fig:dm}).

It is worth noticing that this result in not function of the
X-ray center adopted in our analysis (see discussion in Section~3).
As shown in Fig.~\ref{fig:dm}, moving the center of the circular
annuli used to collect the spectra to the cD galaxy does not change
the mass profile.  Moreover, the presence of a
multi-phase gas could affect our estimate of the gravitational potential.
Gunn \& Thomas (1996) show that, given the same emissivity, the single
phase assumption underestimates the total mass by 20--40 per cent.
Considering that the evidence for a truly multi-phase gas
is weak and the gas is not certainly described from a steady-state
cooling flow model (see, e.g., the discrepancy between the deposition rate
obtained from spatial and spectral deprojection in Section~5.5), 
the correction factors for A1795 would be smaller than these.
Adopting these upper limits and increasing by 40 per cent the mass 
values in the two inner bins do not change the results significantly
(cf. Fig.~\ref{fig:dm_pow}).

\begin{figure}
\psfig{figure=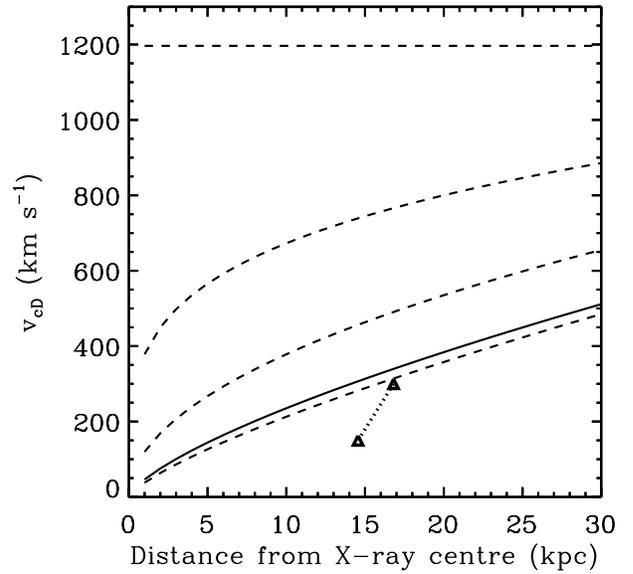,width=.5\textwidth}
\caption[]{Predicted velocity of the cD galaxy under the effect of different
gravitational potentials as a function of the separation from the
X-ray centre assumed to be consistent with the deepest point of the
potential well. The solid line indicates the estimated potential and the
dashed lines are for dark matter density profiles with power law index
of --0.5, --1 (as in NFW), --1.5 (as in Moore et al., 1999) and --2
upwards, respectively. The dotted line shows the range of the values
in the velocity--separation space for angles with respect the line-of-sight
between 0 and 60 degrees. At 0\degr, these values correspond
to the observed peculiar velocity of 150 km s$^{-1}$ and
measured separation between cD galaxy and assumed X-ray centre
of about 9 arcsec. }
\label{fig:motus_cd}
\end{figure}

\begin{figure*}
\hbox{
\psfig{figure=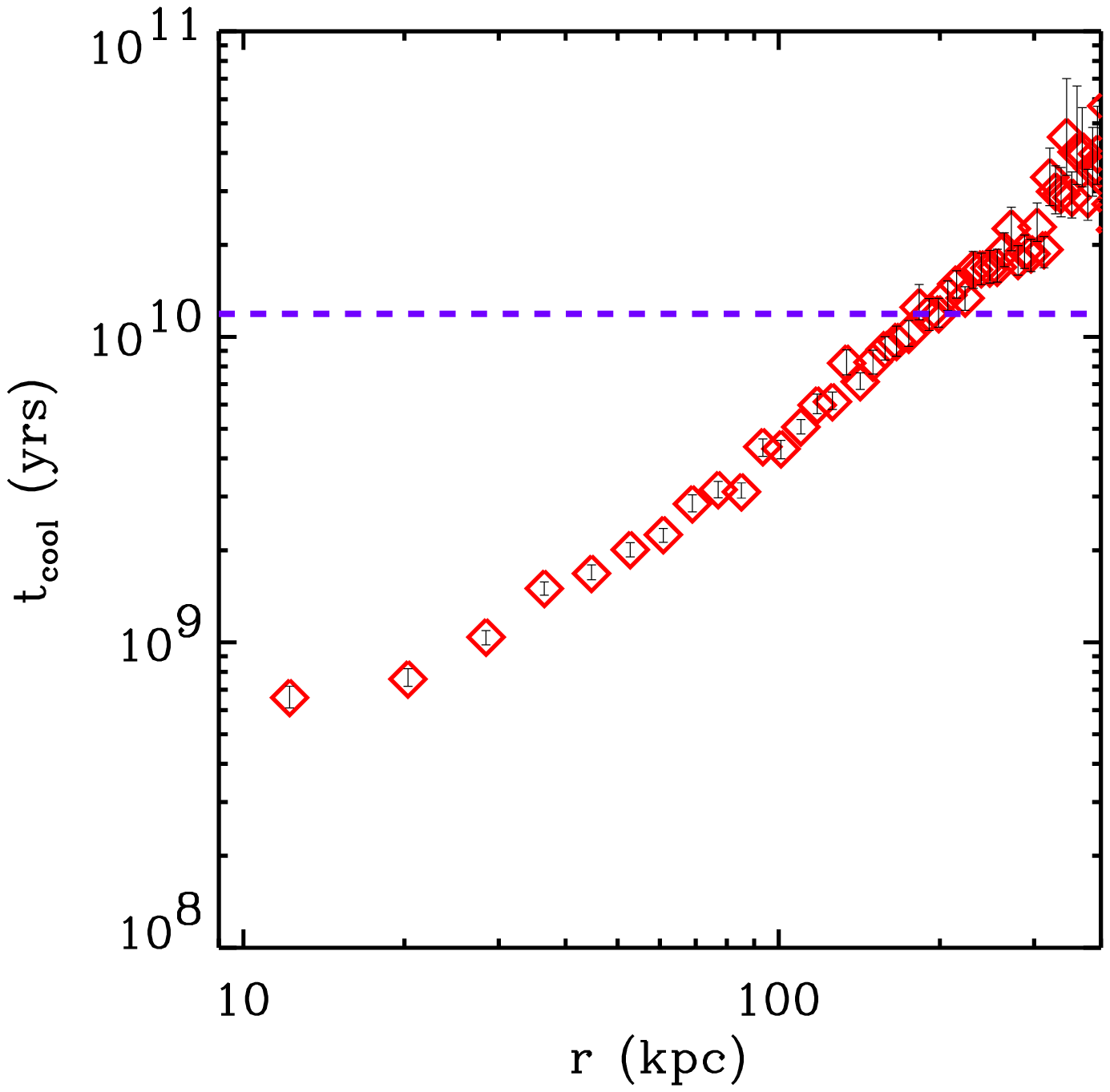,width=.5\textwidth}
\psfig{figure=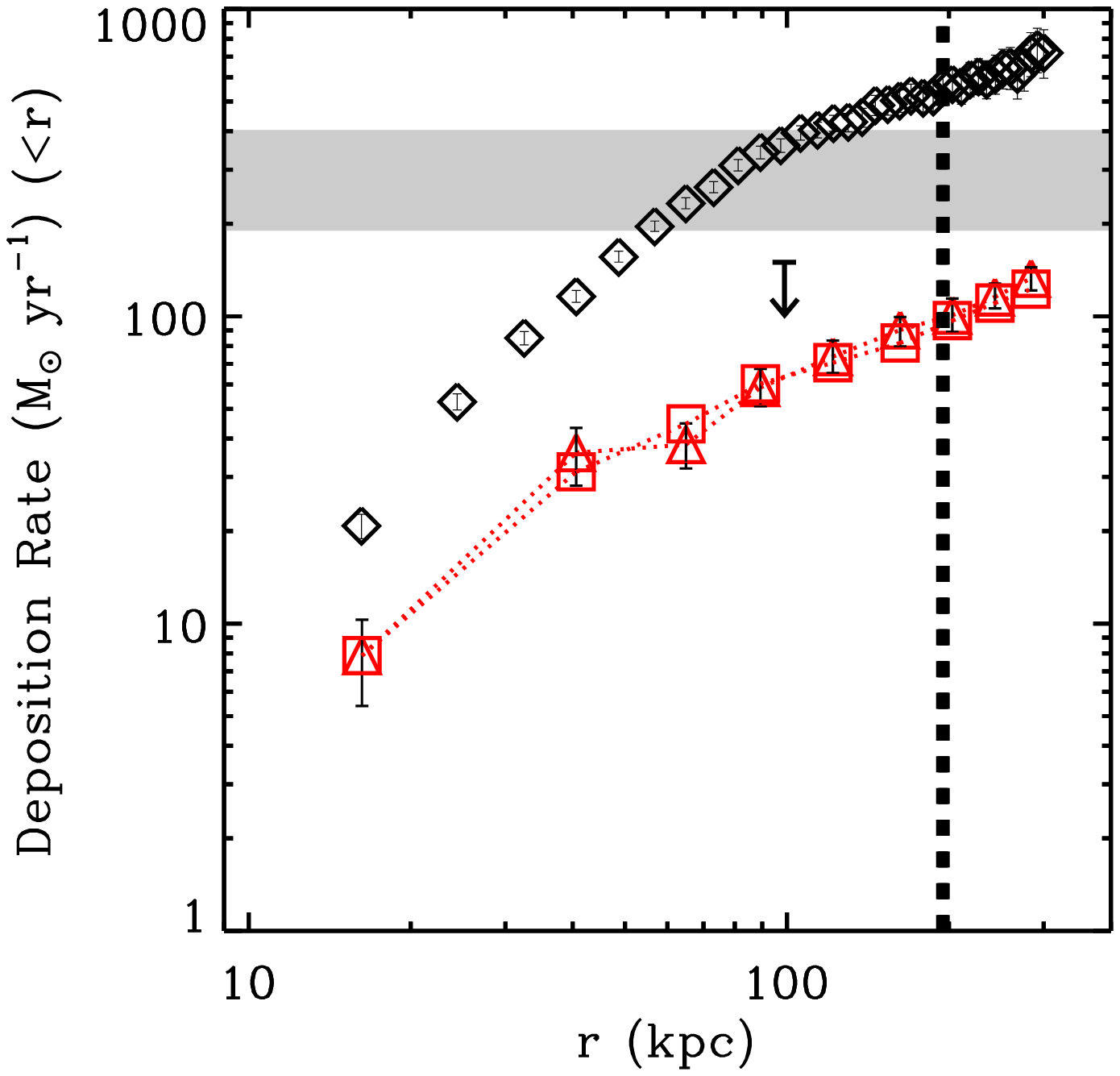,width=.5\textwidth}
}
\caption[]{(Left) {\it Spatial} deprojection results on the cooling time.
The dashed line indicates the age of the universe at the cluster
redshift, $t_{{\rm H}_0}$. Within 10 kpc radius, the cooling time is
about 3.4 $\times 10^8$ yrs.
(Right) {\it Spatial} (diamonds) and {\it spectral}
(triangles: from concentric rings; squares: from annuli)
deprojection results on the integrated
deposition rate, $\dot{M} (<r)$.
The RGS-XMM upper limit is indicated by the downward arrow.
The vertical dashed line corresponds
to the radius of 191 kpc where $t_{\rm cool} = t_{{\rm H}_0}$.
The dashed region shows the 90 per cent confidence level from
\asca analysis (Allen et al. 2001).
} \label{fig:mdot}
\end{figure*}

\subsection{On the dynamical state of the core}

The analysis of the surface brightness distribution in Section~3
suggests that the intracluster medium has an unrelaxed nature that
might introduce a kinetic pressure component in the hydrostatic
equilibrium equation so raising the total effective mass in the
central core. This could possibly explain the flattening of the
gravitating mass profile. Assuming that underlying potential is
described from NFW and $\Delta M$ is the difference with respect to
the observed mass estimates, a bulk motion with velocity (assumed
independent from radial position) of $\sim$ 300 km s$^{-1}$ is
required from the equation $d (v^2 \rho_{\rm gas})/dr = - (G \rho_{\rm
gas} \Delta M) / r^2$. This value is lower than the gas sound speed of
about 800 km s$^{-1}$ and definitely subsonic.

It is worth noting that the shape of the gravitational potential is
consistent with the detected motion of the cD galaxy. Oegerle \& Hill
(1994) measure a cD velocity redshifted relative to the cluster of
$\sim$ 150 km s$^{-1}$. The emission-line velocity map in Hu et al.
(1985) shows that the H$\alpha$/X-ray filament has the same velocity
of the cluster and is blueshifted from the cD velocity by still 150 km
s$^{-1}$ (from recent H$\alpha$ observation, this can be considered
as un upper limit; C. Crawford, priv. comm.). 
In Fig.~\ref{fig:motus_cd}, the cD velocity is estimated as
$v_{\rm cD} = (2 G M_{\rm grav} / r)^{0.5}$, where $G$ is the
gravitational constant and $M_{\rm grav}$ is the total gravitating
mass within the radius $r$ calculated from eq.\ref{eq:mgrav}. Fixing
the gravitational mass at $r$= 100 kpc, we can change the power law
index $\alpha$ and investigate the behaviour of the cD velocity. As
shown in Fig.~\ref{fig:motus_cd}, a cD velocity larger than 400 km
s$^{-1}$ is required for potential well described by a dark matter
density profile steeper than $r^{-1}$. Given the observed velocity
offset along the line of sight, the presence of a density profile flatter 
than a NFW profile is then required. 

If we model the motion of the dominant galaxy as an harmonic oscillation
at the bottom of the potential well (e.g. Lazzati \& Chincarini 1998),
we assume that the cD galaxy passed through the
X-ray cluster centroid at the time zero with maximal velocity and, then, 
underwent an oscillatory motion described from the equations,
$x_{\rm cD} = A \cos(\omega t +\phi), v_{\rm cD} = - A \omega 
\sin(\omega t +\phi)$, where the phase $\phi = -\pi/2$ and
$\omega = (G M_{\rm grav} / r^3)^{0.5} \approx 5 \times 10^{-16}$.
Solving the above two equations, we find that the cD galaxy crossed
the centroid about 1.3 $\times 10^8$ years ago with velocity of $\sim
400$ km s$^{-1}$. Considering our interpretation of the X-ray filament
pointing southward as a cooling wake (see Fabian et al. 2001b for a
discussion), we can use this age as an upper limit on the cooling time
of the densest, coolest phases in the central plasma.

\subsection{On the cooling flow model}

As discussed in Section~4, we cannot conclude that the gas 
in the core of A1795 is multi-phase.  However, this gas
would cool radiatively in about $10^9$ years 
(3.8 $\times 10^8$ yrs in the central 10 kpc radius, with 10th and 90th
percentile of 3.2 and 4.4 $\times 10^8$ yrs, respectively),
approaching the estimated age of the Universe at the cluster redshift
of $1.2 \times 10^{10}$ years at about 200 kpc (Fig.~\ref{fig:mdot}).
This can be considered as an upper limit on the region where
cooling is taking place.
In particular, the physical extension of the central region,
where a X-ray soft component dominates the emission, can be obtained 
from the X-ray colour profile shown in Fig.~\ref{fig:col_rat}, 
where the profile flattens at about 100 kpc. 
This is consistent with the break observed in the
$\dot{M}(r)$ profile discussed below and presented in Fig.~\ref{fig:mdot}.

To maintain the cluster core in pressure equilibrium as 
energy is lost to radiation, and if there is not balancing from 
any heating sources, a flow of material occurs inward and 
deposition of cool gas mass takes place in the central region 
of the cluster. In this {\it cooling flow} scenario (e.g. Fabian 1994),  
the amount of gas deposited within a given radius,
$\dot{M} (<r)$, can be estimated and used to parametrize 
the luminosity originating from the cooling materials.
We measure $\dot{M} (<r)$ in two ways (Fig.~\ref{fig:mdot}): 
in the {\it spatial deprojection},
it is provided from the luminosity associated to the gas that
either (i) cools completely in a shell or (ii) passes the shell
changing temperature under the action of the cluster potential
(cf. White, Jones \& Forman 1997); in the {\it spectral
analysis}, $\dot{M} (<r)$ is estimated as the amount of gas that
cools from the ambient temperature to zero through the modelling
of the spectra accumulated in concentric circular region with
a multi-phase model (Johnstone et al. 1992) absorbed by a column
density intrinsic to the galaxy cluster that is required to reproduce
properly the soft part of the spectra.

The cumulative values of $\dot{M} (<r)$ obtained in the present analysis
(Fig.~\ref{fig:mdot}) enclose the constraints available from the
\asca analysis in Allen et al. (2001), where no spatial resolution
of $\dot{M} (<r)$ was possible.
The deviation between the spatial and spectral results on the 
$\dot{M}$ profile, apart from the central two bins where the assumed
spherical geometry is probably inappropriate given the observed
cooling wake, is due to limitations on the validity of the steady--state 
cooling flow model.
The dynamical scenario in which cooling flows establish 
and evolve considers merging with infalling substructures 
that interrupt the subsonic flow of material. These mergers
are been shown to be responsible for several features
in the surface brightness and temperature distribution 
(see, e.g., Markevitch et al. 2001, Vikhlinin et al. 2001,
Mazzotta et al. 2001) and can be energetically relevant,
thus throwing into question the assumption of steady--state mass accretion
underlying the standard cooling flow model (e.g. Nulsen 1986). 
Therefore, we adopt hereafter the spectral estimates of the 
mass deposition rate as more reliable and use these as 
reference values.

Recent analysis of \xmm data of A1795 (Tamura et al. 2001, 
Molendi \& Pizzolato 2001) does
not show detectable emission from gas cooling below 1--2 keV. Our
lower limit from the deprojection of the best-fit gas temperature in
the central 20 kpc radius is 1.8 keV at 90 per cent confidence level.
When an isobaric cooling flow component is considered to model the \xmm
Reflection Grating Spectrometers (RGS) spectra of the central
region, an upper limit (90 per cent level of confidence) of 150
$M_{\odot}$ yr$^{-1}$ is obtained (Tamura et al. 2001; 
note that these authors estimate that about 80 per cent of the emission 
observed in the RGS spectra comes from regions enclosed within 
$\sim$ 60\arcsec).  This value is
larger than our value of 74 $M_{\odot}$ yr$^{-1}$ (90 per cent
confidence limit of 89 $M_{\odot}$ yr$^{-1}$) measured within 
75 arcsec from the X-ray centre. 
The \chandra-determined cooling flow therefore appears to be consistent 
with the present \xmm RGS constraint. Within the central 200 kpc, 
we estimate a deposition rate of about 100 $M_{\odot}$ yr$^{-1}$ 
($<$ 121 $M_{\odot}$ yr$^{-1}$ at 90 per cent c.l.), consistent with 
the early \asca result of Fabian et al. (1994b) and lower than
the estimate in Allen et al. (2001), where the assumption of an 
isothermal gas in the core (due to the limited spatial resolution of 
\asca) causes an overestimate of the integrated 
deposition rate (up to a factor of $\sim$ 3 in the case of A2390 
discussed in Allen, Ettori \& Fabian 2001).
When a cooling flow model with a low temperature cut-off is considered
(e.g. Tamura et al. 2001), good fits can be obtained for the inner 3 annuli
with lower estimates of $T_{\rm cut-off} \sim$ 1.4 keV. 
With respect to {\it absCF}, this model preserves the same number
of degrees-of-freedom and provides a better $\chi^2$ just for the 
second annulus that encloses a region between 10 and 25 arcsec 
($\Delta \chi^2$= --0.7, 3.3, --0.8 for the inner three annuli, 
respectively, when compared with the results in Table~2).
This is qualitatively in agreement with the results from \xmm EPIC analysis 
in Molendi \& Pizzolato (2001).

Note that we model the cluster emission with {\it absCF},
in which the cool emission is suppressed with an intrinsic absorption
of about $10^{21}$ particle cm$^{-2}$.
If we adopt the same model that describes the RGS spectra and
consider a region of 75 arcsec radius (80 per cent of the emission
observed in the RGS spectra comes from a projected radius of about 
60 arcsec), fixing
the outer thermal component to 6.4 keV with metallicity of 0.4 solar
and the Galactic column density to 3 $\times 10^{20}$ cm$^{-2}$, 
and not including any intrinsic absorption,
we measure a normalization for the cooling flow component
of 140 $M_{\odot}$ yr$^{-1}$ (90 per cent confidence limit range
of 133--152 $M_{\odot}$ yr$^{-1}$) with a considerably worst 
$\chi^2$ of 654 (387 d.o.f.) with respect to 535 (384) obtained
with {\it absCF}. 
On the other hand, a cooling flow model with a low temperature cut-off
provides a better fit ($\chi^2=$ 505, 383 d.o.f.; F-test probability 
of 3 $\times 10^{-3}$) with 
$T_{\rm cut-off} \approx$ 1.4 keV (in the range 1.3--1.6 keV 
at 90 per cent c.l.)
and normalization $\dot{M} \approx$ 324 $M_{\odot}$ yr$^{-1}$
(range at the 90 per cent c.l.: 271--469 $M_{\odot}$ yr$^{-1}$).
The intrinsic absorption discussed above can be interpreted 
as one method to suppress
the line emission from gas below $\sim$ 1.5 keV. Other methods 
and related issues are discussed by Peterson et al. (2001) and 
Fabian et al. (2001a).

\subsection{Distribution of the metals}

The X-ray emitting plasma is about 5 times more massive than the 
stars in galaxies (David et al. 1990, White et al. 1993) 
and, thus, is the major reserve of both baryons and heavy
metals contained in galaxy clusters, since the Iron abundance is typically
between 0.3 and 0.5 times the solar value (Mushotzky \& Loewenstein 1997).
Studies on the correlation between Fe mass and light coming from 
E and S0 galaxies (Arnaud et al. 1992, Renzini 1997) conclude that 
larger amount of iron resides in the intracluster medium than 
inside galaxies and its enrichment originated through releases  
from early-type galaxies.

The processes that preferably enrich the plasma are (i) (proto)galactic 
winds (De Young 1978, Metzler \& Evrard 1994), that occur at early times
and are characterized from supernova (SN) type II ejecta with large abundance
of $\alpha$ elements, and (ii) ram pressure stripping (Gunn \& Gott 1972), 
that takes place on longer time scales due to the continuous accretion of 
fields galaxies in the cluster potential well and produces mostly 
SN Ia ejecta. A negative gradient in the metallicity profiles
is observed preferentially in cooling flow clusters (Allen \& Fabian 1998,
Irwin \& Bregman 2001, 
De Grandi \& Molendi 2001) and can be explained as enrichment of the
inflowing gas by SNe Ia and stellar mass loss in the outer parts of 
the central dominant galaxy (Reisenegger, Miralda-Escud\'e \& Waxman 1996).  

What discriminates between these two main processes is the different
elemental mass yields: SN Ia ejecta tend to be rich in Ni, whereas
SN II ejecta present larger ratio between $\alpha$ elements 
(e.g., O, Mg, Ar, Ca, S, Si) and Fe.
Recent evidence of iron gradients in cluster cores with decreasing 
ratio between $\alpha$ elements and Fe moving inward suggest that, while
the global intracluster metal abundances are consistent with 
SN II ejecta (Mushotzky \& Loewenstein 1997), SN Ia productions
are dominant in the central cluster regions (Ishimaru \& Arimoto 1997,
Fukuzawa et al. 1998, Finoguenov, David \& Ponman 2000, Allen et al. 2001,
Dupke \& Arnaud 2001). 

\begin{figure*}
\hbox{
\psfig{figure=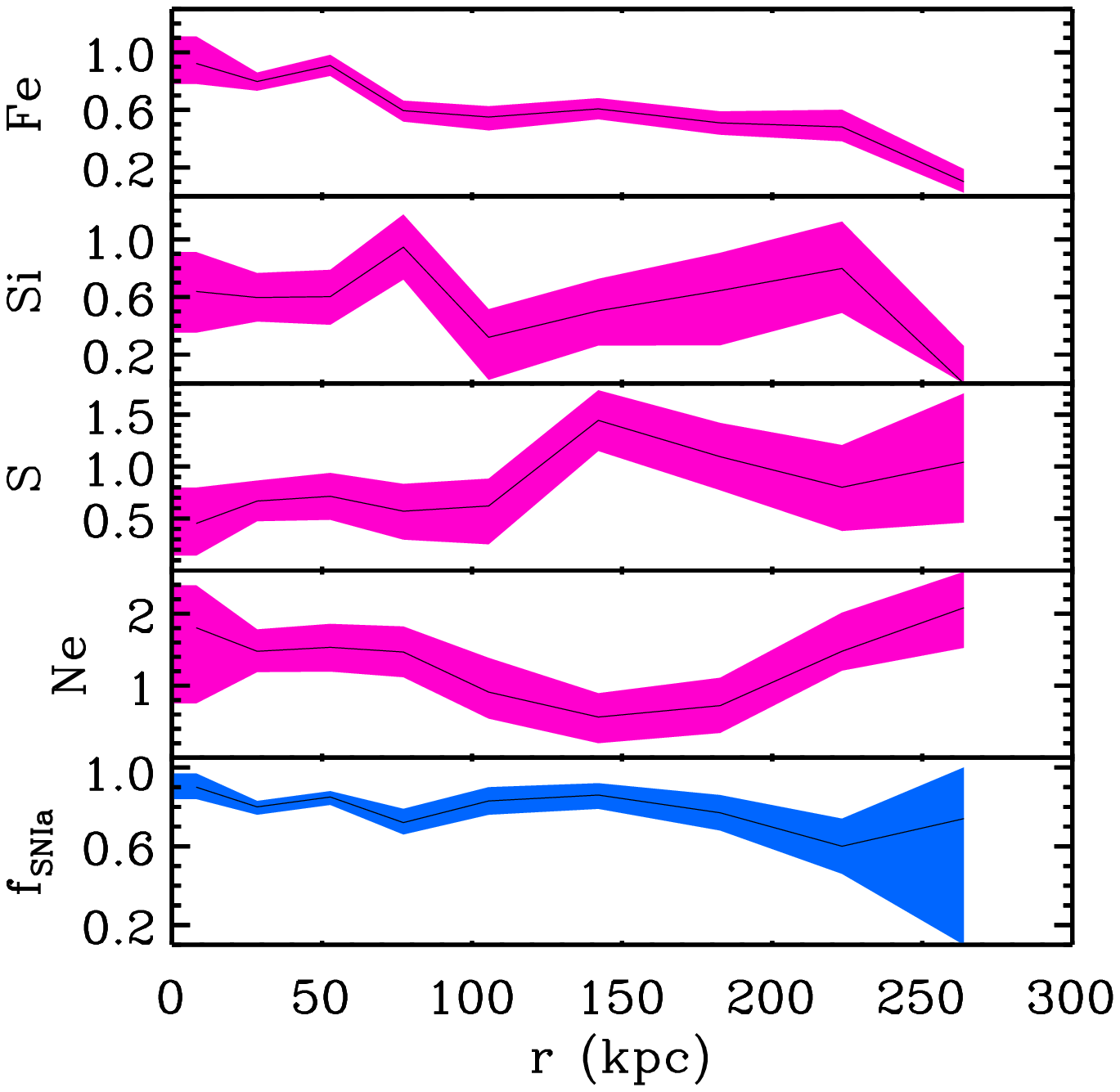,width=.5\textwidth}
\psfig{figure=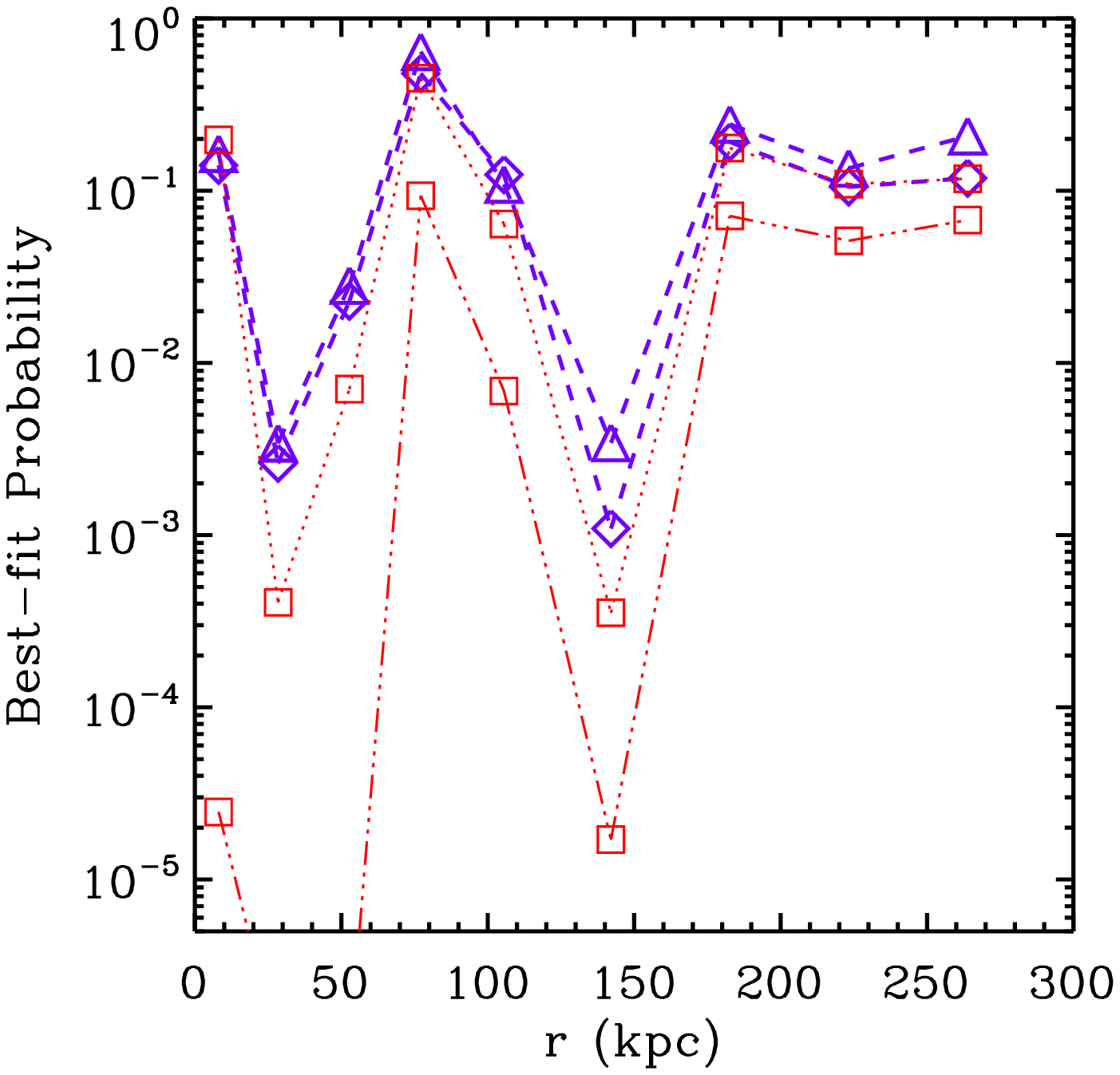,width=.5\textwidth}
} \caption[]{ (Left) Abundance respect to the solar value of Fe, Si, S, Ne.
The O profile is approximately flat around 0.4.
The solar elemental abundance is from Grevesse \& Sauval (1998).
(Note that the results on the azimuthally-averaged metallicity in Table~2 
and Fig.~7 refer at Anders \& Grevesse (1989) for a direct 
comparison with previous work.)
The bottom panel provides the fraction of SNIa that are aspected 
to contribute to the enrichment of the intracluster medium.
(Right) Best-fit Null Hypothesis after using 
single abundance ({\sc Mekal}, {\it diamonds}), multi-abundances
({\sc Vmekal}, {\it triangles}), SNIa metals ratio 
({\it squares} and dotted line), SNII metals ratio
({\it squares} and dash--three--dot line).
} \label{fig:metals}
\end{figure*}

To resolve and quantify the contribution from the single
heavy elements, we fit the annular spectra with
an absorbed {\sc Vmekal} model ({\it abs1TZ} in Table~3)
using the solar elemental abundances  
from Grevesse \& Sauval (1998) that update the 
commonly-used values in Anders \& Grevesse (1989; note that 
the results on the azimuthally-averaged metallicity in Table~2
and Fig.~7 refer at these values for a direct comparison with
previous work) and differ from these mainly in the abundance 
of Iron, Oxygen and Sulphur by a factor of 0.68, 0.79 and 1.32, 
respectively.
We fix Helium to the cosmic abundance value and consider 5 groups of metals:
O, C, N, Na, Mg and Al tied together; Ne; Si; S; Fe tied with 
Ca, Ar and Ni.
In  Fig.~\ref{fig:metals}, we show the profiles for Ne, Si, S and Fe.
We omit to plot the constraints on the value of the O abundance 
that is approximately flat around 0.4 times solar.
In general, these fits show a null-hypothesis probability 
higher than the corresponding single abundance fit
(right panel in Fig.~\ref{fig:metals}).

To estimate the Fe mass fraction originated from SN Ia and SN II ejecta,
we perform a $\chi^2$ minimization at each radial bin of the merit function
\begin{equation}
\chi^2 = \sum_i \frac{ (ratio_i - mod_i)^2 }{err_i^2}
\end{equation}
where $ratio_i$ = [O/Fe, Ne/Fe, S/Fe, Si/Fe], $err_i$ is the 
1 $\sigma$ error on $ratio_i$ propagated from the uncertainties in the
fit and $mod$ is equal to $f \times ratio_{\rm SNIa} + (1-f) 
\times ratio_{\rm SNII}$.
The elemental abundance ratios for SN Ia (0.037, 0.006, 0.585, 0.538 for
O/Fe, Ne/Fe, S/Fe, Si/Fe, respectively) and SN II (3.82, 2.69, 2.29, 
3.53) yields are from Nomoto et al. (1997a, 1997b; for SN Ia yields, we 
refer to deflagration model, W7) and summarized 
in Table~4 in Dupke \& Arnaud (2001).
The radial profile of $f_{\rm SNIa}$ is plotted at the bottom of the 
left panel in Fig.~\ref{fig:metals}.
The fraction in number of the SNe Ia is above 65 per cent everywhere
within 200 kpc.

This is consistent with the higher null-hypothesis probability
that we observe when the abundance ratios are fixed to the 
SN Ia production values instead of SN II values
(right panel in Fig.~\ref{fig:metals}).

A supernova produces 0.74 $M_{\odot}$ of iron when Type Ia (Thielemann,
Nomoto \& Hashimoto 1996), $\sim$ 0.12 $M_{\odot}$ when Type II
(Gibson, Loewenstein \& Mushotzky 1997).
These values imply a total iron mass of $M_{\rm Fe} =
0.74 f_{\rm SNIa} N_{\rm SNIa} +0.12 (1-f_{\rm SNIa}) N_{\rm SNII}$,
where $N_{\rm SNIa}$ and $N_{\rm SNII}$ are the number of 
SN Ia and SN II, respectively, and are estimated comparing
this calculation of the iron mass with what measured from the
observed iron abundance, $Z_{\rm Fe}$, plotted in Fig.~\ref{fig:res_spec}.
In the latter case, $M_{\rm Fe} = A_{\rm Fe} \ y_{\rm Fe, \odot} \ 
Z_{\rm Fe} \ M_{\rm H}$, where $A_{\rm Fe}= 55.8$, $y_{\rm Fe, \odot} =  
3.16 \times 10^{-5}$ from Grevesse \& Sauval (1998) and 
$M_{\rm H} = M_{\rm gas}/1.33$ (we are assuming $\mu=0.6$, $n_{\rm e}
= 1.21 n_{\rm p}$), and rises with radius from $3 \times 10^7 M_{\odot}$
in the inner 10 kpc up to few times $10^9 M_{\odot}$ at $r>100$ kpc.
From the compilation on the rates of SNe Type Ia and II in 
Madau, Della Valle \& Panagia (1998), a conversion factor between 1 and 5
is obtained from $N_{\rm SNII}$ to $N_{\rm SNIa}$.
Using a central value of 3.5 (see also Iwamoto et al. 1999), 
we require a number of SN II
that ranges between $10^8$ and $4 \times 10^{10}$ moving outward.

The amount of the total iron produced in SNIa is then 
$M_{\rm Fe, SNIa} / M_{\rm Fe} \approx \left[1+ 3.5 \times (0.12/0.74)
\times (f_{\rm SNIa}^{-1}-1) \right]^{-1} \approx (0.8-1)$ in the
central 200 kpc.

Assuming that about 80 $M_{\odot}$ of star formation are required 
to generate a supernova Type II (e.g. Thomas \& Fabian 1990), 
we conclude that the cumulative star 
formation that took place during the cluster history is 
$\sim 0.7-2 \times 10^{12} M_{\odot}$ within the central 100-200 kpc,
consistent with the expected amount of gas deposited during the
existence of the cooling flow.
From the $H_{\alpha}$ luminosity of about $10^{42}$ erg s$^{-1}$
coming from the central 12 kpc radius (Cowie et al. 1983),
we infer a present star formation rate (SFR) of about 9 $M_{\odot}$ yr$^{-1}$ 
[Kennicutt 1983; Crawford et al. (1999) indicate a visible SFR of about 
2 $M_{\odot}$ yr$^{-1}$ within 5 kpc-radius from optical observation 
of the central dominant galaxy], 
that has to be compared with an inferred cumulative value 
of about $9 \times 10^9$ $M_{\odot}$ on the same region in exam.
These values can agree with constant star formation rate over the 
last Gyr. 

The deposition of material, star formation activity and residuals 
in form of absorbing dust (as suggested for central regions 
of cooling flow clusters; e.g. Voit \& Donahue 1995, Fabian et al. 1994a,
Allen 2000) can (i) explain the observed excess in the column density  
measurements in the inner 50 kpc of about 2 with respect to the Galactic 
value interpolated from radio maps, and (ii) be responsible for emerging
reprocessed emission in the far infrared.
From IRAS scans over the central 4 arcmin of A1795, Allen et al. (2001)
infer a far-infrared luminosity of $<3.3 \times 10^{44}$ erg s$^{-1}$,
that puts an upper limit on the present SFR of 15 $M_{\odot}$ yr$^{-1}$
(from relation in Kennicutt 1998), 
in agreement with the previous estimates.

Finally, this star formation activity releases energy during 
supernova explosions providing an amount of thermal energy 
per gas particle, $E$ 
\begin{equation}
E = \frac{\eta \ E_{\rm SN} \ N_{\rm SNII}}
{M_{\rm gas}/(\mu m_{\rm p})}
\approx 0.4 \left(\frac{\eta}{0.1}\right) \left(\frac{N_{\rm SNII}}
{10^{10}}\right) \left(\frac{10^{12} M_{\odot}}{M_{\rm gas}}\right) 
{\rm keV}, 
\end{equation} 
where a kinetic energy release by one SN, $E_{\rm SN}$, 
of about 1.3 $\times 10^{51}$ erg (e.g. Iwamoto et al. 1999), 
a factor $\eta$ that represents the efficiency of this kinetic energy
in heating the ICM through galactic winds (and assumed here equal to
10 per cent) and typical values for the inner part of A1795
are adopted.
When we compare $E$ to the thermal energy per particle measured (cf. 
Fig.~\ref{fig:res_spec}), we conclude that processes from SNe explosions
are responsible for about 8 per cent of that energy.

\section{Summary and Conclusions}

\begin{figure}
\psfig{figure=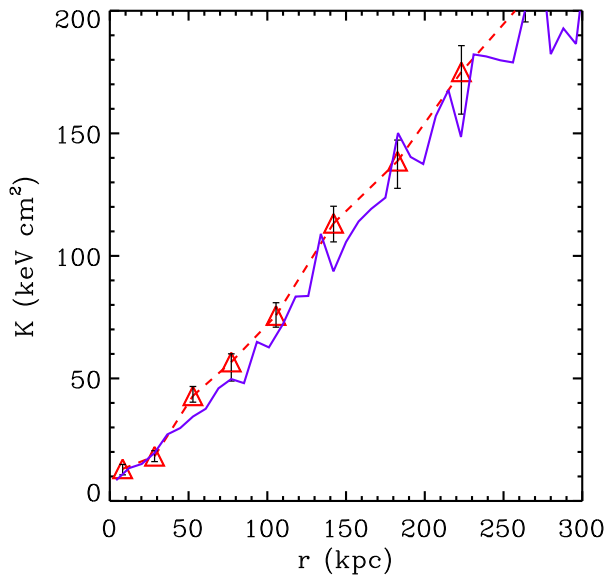,width=.5\textwidth}
\caption[]{Comparison between the {\it spectral} (triangles)
and {\it spatial} (solid line) deprojection results on the
entropy, $K = T_{\rm gas} / n_{\rm gas}^{2/3}$, of the cluster gas.
} \label{fig:entr}
\end{figure}

The entropy of the gas increases monotonically 
moving outwards almost proportionally to the radius,
with a best-fit $\log(K) = -0.09 +0.97\log (r)$ (see Fig.~\ref{fig:entr}).
This behaviour traces the shock heating (more significant at larger
radius) at which the intracluster gas has been subjected during its history
of accretion in the gravitational potential well, whereas the 
efficiency of the cooling has removed the central gas at lower entropy
(cf. Tozzi \& Norman 2001).

Together with the estimated cooling time due to energy loss by radiation 
of less than $10^9$ years and a decrease by a factor of 3 in the gas
temperature, this seems to indicate that the central 200 kpc are
undergoing significant cooling with a gas mass deposition 
rate of about 100 $M_{\odot}$ yr$^{-1}$ in the absence of any
heating process.
This would require an intrinsic absorption of about $10^{21}$ cm$^{-2}$. 
However, a simple single phase model with absorption in excess of 
the Galactic value by a factor of 2, within the inner 50 kpc,
is consistent with the observed spectra.

The gravitational potential follows a NFW profile between 60 and 300 
kpc, becoming flatter within 60 kpc with a power law index of about $-0.6$
once a power law description is adopted for the total matter density.
The shape of the potential is in agreement with the motion of the
central dominant galaxy and suggests that the central cluster region 
is not relaxed.

\begin{figure}
\psfig{figure=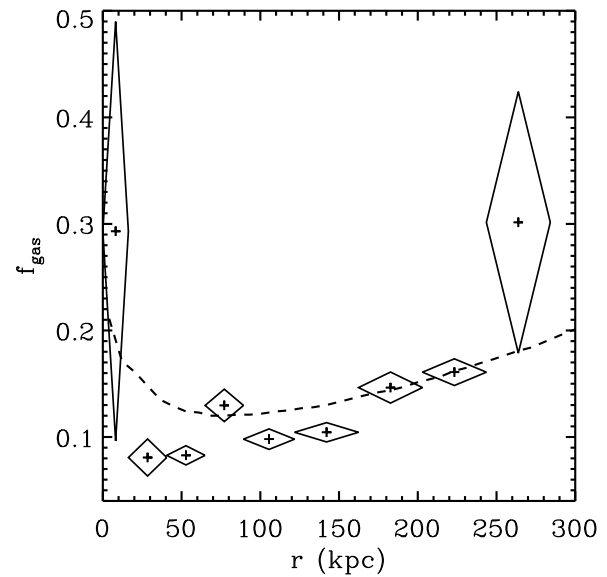,width=.5\textwidth}
\caption[]{Comparison between the {\it spectral} (diamonds with error
bars) and {\it spatial} (dashed line) deprojection
results on the gas mass fraction.
} \label{fig:fgas}
\end{figure}

This dark matter distribution, combined with the integrated gas mass
within 300 kpc, provides a gas mass fraction ranging between 10 and 20 
per cent (Fig.~\ref{fig:fgas}), in a good agreement with the 
estimate of $0.16\pm0.01$ at 500 kpc from \pspc brightness profile 
and \asca temperature in Ettori \& Fabian (1999) and 
completely consistent with that generally observed in galaxy clusters
(Evrard 1997, Ettori \& Fabian 1999, Mohr et al. 1999). 
 
Finally, the present \chandra observation of the core of A1795
allows us to resolve radially the contribution from Oxygen, Neon, 
Sulphur, Silicon and Iron to the plasma metallicity and, 
consequently, to estimate between 80 and 100 per 
cent the amount of Iron produced in supernova Type Ia events
within the central 200 kpc.

\section*{ACKNOWLEDGEMENTS} 
We thank all the staff members involved in the \chandra project for 
such an excellent performance of the instruments.
ACF is grateful to NASA for the opportunity to participate
to the project as an InterDisciplinary Scientist.
ACF and SWA thank the Royal Society for support.
We thank the referee for a careful reading of the manuscript 
and for several suggestions that improved the presentation of this work.

\end{document}